\documentclass[a4paper,11pt]{article}
\pdfoutput=1
\usepackage{authblk}
\usepackage{abstract}
\usepackage[footnotesize]{caption}
\usepackage{subcaption}
\usepackage{cite}
\usepackage{amsmath}
\usepackage{amssymb}
\usepackage{graphicx} 
\usepackage[latin1]{inputenc}
\usepackage{mathrsfs}
\usepackage[dvipsnames]{xcolor}
\usepackage{hyperref}
\usepackage[hmarginratio=1:1,top=32mm,columnsep=20pt]{geometry}
\usepackage{paralist}
\usepackage{multirow}
\usepackage{multicol}
\usepackage[english]{babel}
\usepackage{blindtext}
\newcommand{\bea}{\begin{eqnarray}}
\newcommand{\eea}{\end{eqnarray}}
\newcommand{\be}{\begin{equation}}
\newcommand{\ee}{\end{equation}}
\newcommand{\ba}{\begin{array}}
\newcommand{\ea}{\end{array}}

\def\gsim{\mathrel{\rlap{\lower4pt\hbox{\hskip1pt$\sim$}}
    \raise1pt\hbox{$>$}}}

\renewcommand\Affilfont{\normalsize\itshape}

\title{\fontsize{16pt}{10pt}\selectfont
	\textbf{Low scale type II seesaw: Present constraints \\[1mm] and prospects for displaced vertex searches}
	}
\author[1]{Stefan~Antusch\thanks{\texttt{stefan.antusch@unibas.ch}}}
\author[2]{Oliver~Fischer\thanks{\texttt{oliver.fischer@kit.edu}}}

\author[1]{A.~Hammad\thanks{\texttt{ahmed.hammad@unibas.ch}}}
\author[1]{Christiane Scherb\thanks{\texttt{christiane.scherb@stud.unibas.ch }}}

\affil[1]{\Affilfont Department of Physics, University of Basel, \authorcr 
 		  \Affilfont Klingelbergstr.\ 82, CH-4056 Basel, Switzerland
 		   \authorcr\mbox{}}
\affil[2]{\Affilfont Institute for Nuclear Physics, Karlsruhe Institute of Technology
           \authorcr 
 		  \Affilfont  Hermann-von-Helmholtz-Platz 1, D-76344 Eggenstein-Leopoldshafen, Germany
 		   \authorcr\mbox{}} 		   
\date{}

\begin{document}
\maketitle

\setlength{\absleftindent}{50pt}
\setlength{\absrightindent}{50pt}
\vspace{-15pt}
\begin{abstract}
The type II seesaw mechanism is an attractive way to generate the observed light neutrino masses. It postulates a SU(2)$_\mathrm{L}$-triplet scalar field, which develops an induced vacuum expectation value after electroweak symmetry breaking, giving masses to the neutrinos via its couplings to the lepton SU(2)$_\mathrm{L}$-doublets. When the components of the triplet field have masses around the electroweak scale, the model features a rich phenomenology. We discuss the current allowed parameter space of the minimal low scale type II seesaw model, taking into account all relevant constraints, including charged lepton flavour violation as well as collider searches. 
We point out that the symmetry protected low scale type II seesaw scenario, where an approximate ``lepton number''-like symmetry suppresses the Yukawa couplings of the triplet to the lepton doublets, is still largely untested by the current LHC results. 
In part of this parameter space the triplet components can be long-lived, potentially leading to a characteristic displaced vertex signature where the doubly-charged component decays into same-sign charged leptons. By performing a detailed analysis at the reconstructed level we find that already at the current run of the LHC a discovery would be possible for the considered parameter point,
via dedicated searches for displaced vertex signatures.
The discovery prospects are further improved at the HL-LHC and the FCC-hh/SppC. 

\end{abstract}
\hrulefill
\vspace{1cm}

\noindent

\section{Introduction}
The Standard Model (SM) of elementary particles is successfully describing a plethora of observed phenomena at many different energy scales.
However, the observation of neutrino oscillations \cite{Ahmad:2002jz,Fukuda:1998mi} is evidence that at least two of the neutrinos are massive. Since the SM cannot account for these masses in a renormalizable way, this calls for physics beyond the SM (BSM).
An attractive possibility for generating the masses for the neutrino degrees of freedom of the SM consists in adding a scalar SU(2)$_\mathrm{L}$-triplet field (a ``triplet Higgs field'') to the scalar sector of the theory, which obtains an induced vacuum expectation value $v_T$ after electroweak symmetry breaking, giving masses to the neutrinos via its couplings to two lepton SU(2)$_\mathrm{L}$-doublets. This mechanism for neutrino mass generation is often referred to as the type-II seesaw mechanism \cite{Konetschny:1977bn,Magg:1980ut,Schechter:1980gr,Cheng:1980qt,Mohapatra:1980yp,Lazarides:1980nt}.

In particular the ``low scale'' version of the type II seesaw mechanism, where the components of the triplet field have masses around the electroweak scale (or TeV scale), has implications for various well known observables at different energy scales, see e.g. \cite{Abada:2007ux,Dev:2018sel}. It may be embedded for instance in left-right symmetric extensions of the SM, with additional interesting phenomenology at the LHC, cf.\ refs.\ \cite{Maiezza:2015lza,Nemevsek:2016enw}, or studied in its minimal version with only one triplet Higgs added to the SM. Regarding the triplet Higgs field, its doubly charged component is of particular importance for phenomenology, since it can decay into a pair of same-sign charged leptons via the above mentioned lepton number violating Yukawa coupling (matrix) $Y_\Delta$ of the triplet to the lepton SU(2)$_\mathrm{L}$-doublets. Detailed phenomenological studies of such signatures have been conducted for the LHC, e.g.\ in refs.\ \cite{Akeroyd:2005gt,Perez:2008ha,Melfo:2011nx,Foot:1988aq}, and also for a 100 TeV proton-proton collider in ref.~\cite{Du:2018eaw}.

Searches for prompt decays to same-sign lepton pairs and pair-produced doubly charged Higgs bosons have been performed at the LHC (for the different center-of-mass energies) \cite{ATLAS:2012hi,CMS:2011sqa,ATLAS:2014kca,CMS:2016cpz,Aaboud:2017qph,CMS:2017pet}, and similar analyses exist for LEP \cite{Abbiendi:2001cr,Abdallah:2002qj,Achard:2003mv}, and at the Tevatron \cite{Acosta:2004uj,Aaltonen:2008ip,Abazov:2008ab,Abazov:2011xx}. Searches for same-sign $W$ boson pairs have recently been performed at LHC in ref.~\cite{Aaboud:2018qcu}. 
Without any significant excess of events, the LHC analyses mentioned above presently provide stringent constraints from direct searches, which require the masses of the doubly charged scalars to be above $\sim 600$ GeV (for the part of parameter space where $Y_\Delta$ is not too small).
Moreover, searches at future lepton colliders could have the potential to discover doubly charged scalars with masses $\sim 1$~TeV, provided the center-of-mass energy is 3~TeV, as discussed in ref.\ \cite{Agrawal:2018pci}.

The possibility that the scalar particles do not decay promptly, but can be rather long lived, has important consequences for LHC searches: While the above mentioned strong constraints from prompt same-sign charged leptons can no longer be applied, one might consider them as heavy Stable Charged Particles (HSCPs) if their lifetime is sufficiently long for them to pass through the relevant parts of the detector, i.e.\ the muon system (or the tracker). The corresponding signature would be, among others, a characteristic energy deposition in the different subdetectors. Searches for HSCPs have been performed by ATLAS \cite{Aad:2013pqd,Aad:2015oga} and CMS \cite{Khachatryan:2016sfv}. When the decays of a long lived particle are non-prompt but occur inside the detector, one might also search for the displaced secondary vertices. This possibility has recently been discussed in ref.\ \cite{Dev:2018kpa}, where it has been claimed that the high-luminosity (HL) LHC can probe a broad part of the parameter space via such displaced vertex searches, restricted however severely by the HSCP constraints. 

In this paper we discuss the current allowed parameter space of the minimal low scale type II seesaw model, taking into account all relevant constraints, including charged lepton flavour violation as well as various (prompt and non-prompt) collider searches. We calculate carefully the constraints from the prompt searches, taking into account only the simulated events which satisfy the ``promptness'' criteria applied in the experimental analyses. Reconsidering constraints from HSCP searches, we find that the existing analyses cannot be applied to the triplet components of the minimal type II seesaw because their lifetimes are not large enough to pass through a sufficient part of the detector. Finally, for the displaced vertex signature, we perform a detailed analysis at the reconstructed level, for a selected benchmark point. We find that already at the current run of the LHC, a discovery would be possible for the considered parameter point. At a future collider with higher center-of-mass energy like the FCC-hh/SppC \cite{Golling:2016gvc,Tang:2015qga}, the larger Lorentz factors and larger luminosities would further enhance the sensitivity of these displaced vertex searches.

\section{The minimal type II seesaw extension of the Standard Model}
In the minimal type-II seesaw model the scalar sector consists of the SM scalar $\Phi\sim (1,2,\frac{1}{2})$ and an additional triplet scalar field $\Delta\sim(1,3,2)$. Their matrix representation is given by:
\begin{equation}
	\Phi =  \left( \begin{array}{c}\Phi^+\\\Phi^0 \end{array} \right)   \text{ and } \Delta = \begin{pmatrix}
	\frac{\Delta^+}{\sqrt{2}}&\Delta^{++}\\
	\Delta^0&-\frac{\Delta^+}{\sqrt{2}}
	\end{pmatrix} .
\end{equation}
The $SU(3)_C\times SU(2)_L\times U(1)_Y$ invariant Lagrangian for this scalar sector is
\begin{align}
\mathcal{L} =\, & (D_{\mu}\Phi)^{\dagger}(D^{\mu}\Phi)+Tr((D_{\mu}\Delta)^{\dagger}(D^{\mu}\Delta))-V(\Phi,\Delta) -\mathcal{L}_{Yukawa}
\label{Lagrangian} 
\end{align}
with the covariant derivaties
\begin{align}
&D_{\mu}\Phi=\partial_{\mu}\Phi+igT^aW^a_{\mu}\Phi+i\frac{g'}{2}B_{\mu}\Phi \\
&D_{\mu}\Delta=\partial_{\mu}\Delta+ig[T^a_{\mu},\Delta]+i\frac{g'}{2}B_{\mu}\Delta,
\label{derivaties}
\end{align}
the scalar potential
\begin{align} 
V(\Phi,\Delta) =\, & \mu^2 \Phi^{\dagger}\Phi - M^2_T Tr(\Delta^{\dagger}\Delta)  - \frac{\lambda}{4} |\Phi^{\dagger}\Phi|^2 \nonumber \\
&- \lambda_{HT}\Phi^{\dagger}\Phi Tr(\Delta^{\dagger}\Delta) -\lambda_T (Tr(\Delta^{\dagger}\Delta))^2 \nonumber \\
&-{\lambda'}_T Tr((\Delta^{\dagger}\Delta)^2) -{\lambda'}_{HT} \Phi^{\dagger}\Delta\Delta^{\dagger}\Phi  \nonumber \\
&-(\kappa\Phi^{\top}i\sigma^2\Delta^{\dagger}\Phi +h.c.)
\label{potential} 
\end{align} 
and the new Yukawa terms
\begin{equation}
\mathcal{L}_{Y_\Delta}=Y_\Delta \bar{\ell^c}i\sigma^2\Delta \ell + H.c.\:.
\label{yukawa}
\end{equation}
After electroweak symmetry breaking both scalar fields acquire their vacuum expectation values (VEVs)
\begin{equation}
\langle\Phi\rangle =\frac{1}{\sqrt{2}}\left( \begin{array}{c}0\\v \end{array} \right) \text{ and } \langle\Delta\rangle = \frac{1}{\sqrt{2}}\begin{pmatrix}
0&0\\
v_T&0.
\end{pmatrix} ,
\end{equation}
where (as we will see later) $v_T \ll v$.
Evolving the scalar fields around their VEVs and minimizing the potential leads to seven physical massive eigenstates: $H^{\pm\pm},H^{\pm},h,H,A$. The three massless Goldstone bosons $G^{\pm}$ and $G^0$ are absorbed by the SM gauge bosons $W^{\pm}$ and $Z$. The masses for the physical Higgs bosons are
\begin{equation}
m^2_{H^{\pm\pm}} =  \frac{\kappa v^2}{\sqrt{2} v_T} + \frac{{\lambda'}_{HT} v^2}{2} + {\lambda'}_T v_T^2
\label{mTPP}
\end{equation}

\begin{equation}
m^2_{H^{\pm}} = \frac{\kappa v^2}{\sqrt{2} v_T} +\frac{{\lambda'}_{HT} v^2}{4} +\frac{{\lambda'}_{HT} v_T^2}{2}+  \sqrt{2}\kappa v_T
\label{mTP}
\end{equation}

\begin{equation}
m^2_h = \frac{1}{2}(A+C-\sqrt{(A-C)^2+4B^2})
\label{SMhiggsmass}
\end{equation}

\begin{equation}
m^2_H = \frac{1}{2}(A+C+\sqrt{(A-C)^2+4B^2}),
\end{equation}
with $A = -\frac{\lambda}{2} v^2$, $B = -(\lambda_{HT}+{\lambda'}_{HT}) v_T v - \sqrt{2}\kappa v$ and $C =\frac{\kappa v^2}{\sqrt{2} v_T} -2(\lambda_T+{\lambda'}_T)v_T^2$, and

\begin{equation}
m^2_{A^0} = \frac{\kappa v^2}{\sqrt{2}v_T}  + 2\sqrt{2}\kappa v_T.
\end{equation}

\paragraph{Physical masses and parameter space:}
The scalar potential and the new Yukawa term contain the following parameters: five coupling parameters $\lambda$, $\lambda_{HT}$, $\lambda_T$, ${\lambda'}_T$ and ${\lambda'}_{HT}$, two mass parameters $\mu$ and $M_T$, the seesaw parameter $\kappa$ (with mass dimension $= 1$), the VEVs $v$ and $v_T$ and the new Yukawa couplings matrix $(Y_\Delta)_{ij}$. 
The tadpole equations allow us to express $\mu$ and $M_T$ in terms of the couplings and $\kappa$:
 \begin{align}
&\mu^2 = -\sqrt{2}\kappa v_T+\frac{1}{2}(\lambda_{HT}+{\lambda'}_{HT})v_T+\frac{1}{4}\lambda v^2 \:,
\label{tadpolemu}
\end{align}
\vspace{-0.75cm}
\begin{align}	
&M_T^2 = -(\lambda_T+{\lambda'}_T)v_T^2-\frac{1}{2}(\lambda_{HT}+{\lambda'}_{HT})v^2+\frac{\kappa v^2}{\sqrt{2}v_T}.
\label{tadpolemt}
\end{align} 
In the following we fix the VEV $v$ to the SM value $v \approx 246$ GeV. By solving the tadpole equations and 
taking the leading order in $\lambda^{(')}_{HT} v_T / \kappa$ we obtain for $v_T$ the relation
 \begin{equation}
v_T = \frac{\kappa v^2}{\sqrt{2}M_T^2}
\label{v_T} \:.
\end{equation}
Furthermore, we chose $h$ to play the role of the SM Higgs boson (with the requirement that $m_h < m_H$), and we fix $\lambda$ such that $m_h \sim 125$ GeV. Neglecting the terms in eq.~\eqref{SMhiggsmass} that are proportional to the triplet VEV $v_T$, we thus use the SM value for $\lambda$.

The contributions from the couplings $\lambda_T$ and ${\lambda'}_T$ to all the mass terms are suppressed by the triplet VEV $v_T$, and we will neglect this contribution in the following discussion. For definiteness, in our analyses we will fix the couplings in the following way: $\lambda_T =0.1$ and ${\lambda'}_T=0.2$. 
The masses of the singly charged scalar $H^{\pm}$ and the doubly charged scalar $H^{\pm\pm}$ depend only via the first term in eqs.~\eqref{mTP} and \eqref{mTPP} on ${\lambda'}_{HT}$, respectively. Their masses are fixed to the same scale by $\kappa v^2/v_T$, with a mass splitting $m_{H^\pm}-m_{H^{\pm\pm}} = -\lambda'_{HT}v^2/4 + \sqrt{2}\kappa v_T$, such that $m_{H^\pm}$ and $m_{H^{\pm\pm}}$ are effectively free parameters. 

In the following, we allow in most cases for a non-zero $\lambda'_{HT} $, but we keep $\lambda'_{HT} < 0$ such that $H^{\pm\pm}$ is the lightest of the new scalars. The reason for this choice is that when we discuss potentially long-lived $H^{\pm\pm}$ (cf.\ section \ref{sec:long-lived}) it avoids additional decay modes, but allows to have $m_{H^{\pm\pm}}$ somewhat below $m_h$. Only for illustrating some of the phenomenological constraints we will make the simplifying assumption that $\lambda'_{HT}=0$, which leads to nearly degenerate masses for all extra scalars (controlled by the parameters $\lambda_{HT}$ and $\kappa$).
We use Sarah \cite{Staub:2013tta} and Spheno \cite{Porod:2003um,Porod:2011nf} for the evaluation of the model parameters and for the numerical calculation of the constraints from non-collider experiments in section 3.

\section{Constraints from non-collider experiments}

\paragraph{Neutrino masses:}
In the type-II seesaw model the active neutrinos acquire masses after electroweak symmetry breaking via the contributions from the new Yukawa term, yielding
\begin{equation}
m_{\nu} = Y_\Delta \sqrt{2}v_T = Y_\Delta \frac{\kappa v^2}{M_T^2}.
\label{neutrinomassterm}
\end{equation} 
It is referred to as a ``seesaw'' model, because the light neutrino masses are inversely proportional to the triplet mass (squared).  

Via eq.~\ref{neutrinomassterm}, the observed neutrino masses constrain the model parameters $(Y_{\Delta})_{ij}$ and $v_T$. 
For a given neutrino mass ordering the Yukawa couplings can be obtained via
\begin{equation}
(Y_{\Delta})_{ij}= \frac{1}{\sqrt{2}v_T}U_{\mathrm{PMNS}}^{\dagger}m_{\nu}^{diag}U_{\mathrm{PMNS}}
\label{UPMNS},
\end{equation}    
where $U_{\mathrm{PMNS}}$ is the  Pontecorvo-Maki-Nakagawa-Sakata matrix. 
In the following, normal hierarchy is assumed and best fit values for $U_{\mathrm{PMNS}}$ are used from \cite{Esteban:2016qun,nufit} (with the additional assumption of the Majorana phase being zero). 
Eq.~\eqref{UPMNS} thus fixes the Yukawa couplings $(Y_{\Delta})_{ij}$ for our choice of assumptions.

\paragraph{Constraints on $v_T$:}
From electroweak precision measurements the $\rho$ parameter is measured to be $\rho \simeq 1.00037\pm0.00023$ \cite{Patrignani:2016xqp}. In the model $\rho$ can be written as
\begin{equation}
\rho = \frac{1+\frac{2v_T^2}{v^2}}{1+\frac{4v_T^2}{v^2}},
\label{modelrho}
\end{equation}
which leads to an upper bound for the triplet VEV $v_T \lesssim 2.1$ GeV. 

\paragraph{Z width:} For a doubly charged mass, $m_ {H^{\pm\pm}} < \frac{m_Z} { 2} $ a new on-shell decay mode $Z\to H^{\pm\pm}H^{\mp\mp}$ is allowed. The LEP experiment constrained the allowed decay width of the $Z$ boson into non-SM particles to be below $2$ MeV at $95\%$ CL, which implies the lower limit on the mass $m_{H^{\pm\pm}} > 42.9$ GeV \cite{Kanemura:2013vxa}.

\paragraph{Lepton flavor violating processes:}
In the type II seesaw model, lepton flavor violating (LFV) processes $\tau\to\bar{l_i}l_jl_k$ and $\mu\to\bar{e}ee$ can be mediated at tree level via $H^{\pm\pm}$ exchange. The contribution of the doubly charged scalars to the LFV branching ratio  $BR(l_i\to l_kl_ml_n)$ is given by\cite{Kakizaki:2003jk}:
\begin{align}
BR(l_i\to \bar l_kl_ml_n) = \frac{|(Y_\Delta)_{mn}(Y_\Delta)_{ki}|^2}{64G^2_f m^4_{H^{\pm\pm}}}\:.
\end{align} 
The most stringent bound arises from $\mu\to \bar{e}ee$ with $BR(\mu\to \bar{e}ee) < 1.0 \times 10^{-12}$ from the SINDRUM experiment \cite{Bellgardt:1987du}. 
Since the Yukawa couplings are inversely proportional to the triplet VEV, the experimental bounds constitute (for our choice of PMNS parameters and neutrino mass spectrum) a lower limit for $v_T$, e.g.\ $v_T > 8.8\times10^{-9}$ GeV, $v_T > 5.1\times10^{-9}$ GeV and $v_T > 3.1\times10^{-9}$ GeV for masses  $m_{H^{\pm\pm}} = 150$ GeV, $m_{H^{\pm\pm}} = 300$ GeV and $m_{H^{\pm\pm}} = 600$ GeV respectively.  
	
Also the lepton flavor violating process $\mu\to e\gamma$ receives contributions from loops with virtual $H^+,H^-,\nu_\alpha$ or $H^{++},H^{--}l_\alpha$, where the appearing couplings to the new scalars are inversely proportional to the triplet VEV.  
The MEG collaboration states the currently most stringent upper bound of $BR(\mu \to e \gamma) < 4.2\times 10^{-13}$ \cite{TheMEG:2016wtm} on the branching ratio of this process, which translates (for our choice of PMNS parameters and neutrino mass spectrum) into a lower limit of the triplet VEV of, e.g., $v_T > 4.8\times10^{-9}$~GeV, $v_T > 2.6\times10^{-9}$~GeV and $v_T > 1.6\times10^{-9}$~GeV for masses $m_{H^{\pm\pm}} = 150$~GeV, $m_{H^{\pm\pm}} = 300$~GeV and $m_{H^{\pm\pm}} = 600$~GeV respectively. 
A discussion of the dependence of the LFV constraints on the PMNS parameters and neutrino mass spectrum can be found e.g.\ in ref.~\cite{Akeroyd:2009nu,Dev:2017ouk}.

\paragraph{The anomalous magnetic moment of the muon:}
The anomalous magnetic moment of the muon was measured very precisely by the Muon g-2 collaboration \cite{Bennett:2006fi}:
	\begin{align*}
	a^{exp}_\mu = 11659208.0(6.3)\times 10^{-10} \:.
	\end{align*}
The result deviates by about three standard deviations from the SM predicted value, given by \cite{Patrignani:2016xqp}:
	\begin{align*}
	a^{SM}_\mu = 11659183\times 10^{-10} \:.
	\end{align*}
	
The type II seesaw model modifies the theory prediction for this amplitude: 
At one loop level the amplitude receives new contributions from both $H^{\pm\pm}$ and $H^\pm$ as 
	\begin{align*}
	\delta a_\mu (H^{\pm\pm}) = - \frac{2\left| (Y_\Delta)_{ij}(Y_\Delta)_{ij}\right|^2m_{\mu}^2}{12\pi^2 m^2_{H^{\pm\pm}}}\:,
	\end{align*}
	\begin{align*}
	\delta a_\mu (H^{\pm}) = - \frac{2\left|(Y_\Delta)_{ij}(Y_\Delta)_{ij}\right|^2 m_{\mu}^2}{96\pi^2 m^2_{H^{\pm}}}\:.
	\end{align*} 	
We notice that, in principle, the modified theory prediction could explain the observed value of $a_\mu$ for some range of the triplet mass and $\upsilon_T \lesssim 10^{-10}$ GeV. This region is, however, already excluded by the LFV experiments.

\section{Signatures from doubly charged scalars at the LHC}
\begin{figure}
\centering
\includegraphics[width=0.65\linewidth]{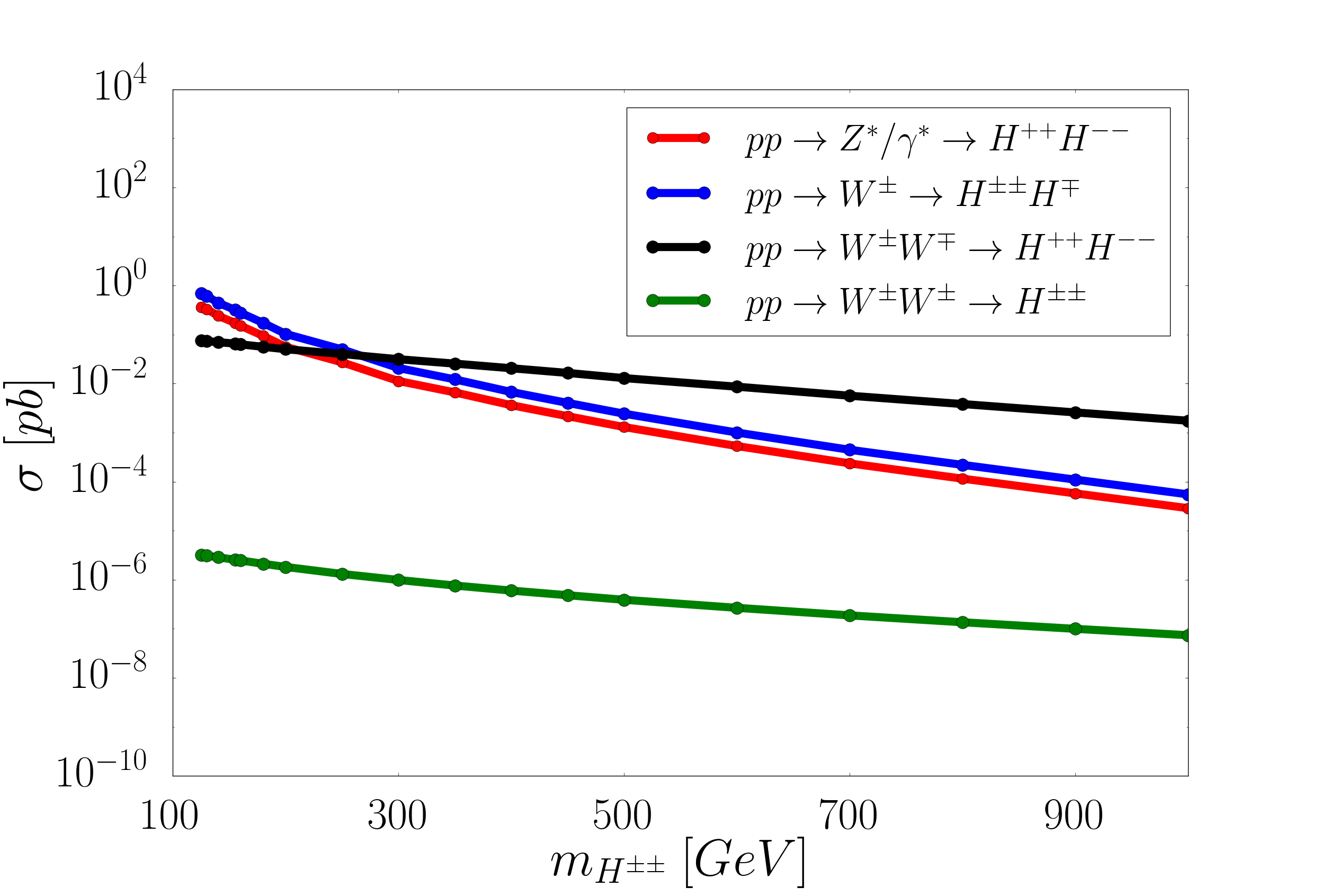}
\caption{Production cross section for the dominant production channels at the LHC with $\sqrt{s}=13$~TeV, the example values $\upsilon_T = 0.1$ GeV for the triplet Higgs vev and $\lambda'_{HT} = 0$.  }
\label{cross_section} 
\end{figure}

\begin{figure}
\centering
\includegraphics[width=0.45\linewidth]{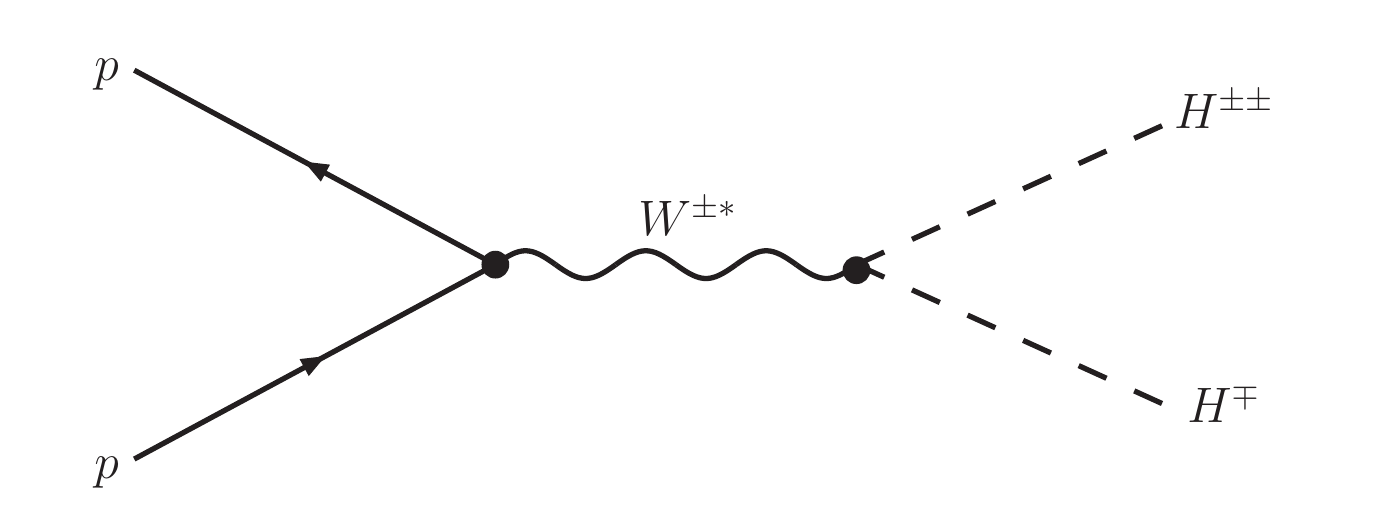}~~
\includegraphics[width=0.4\linewidth]{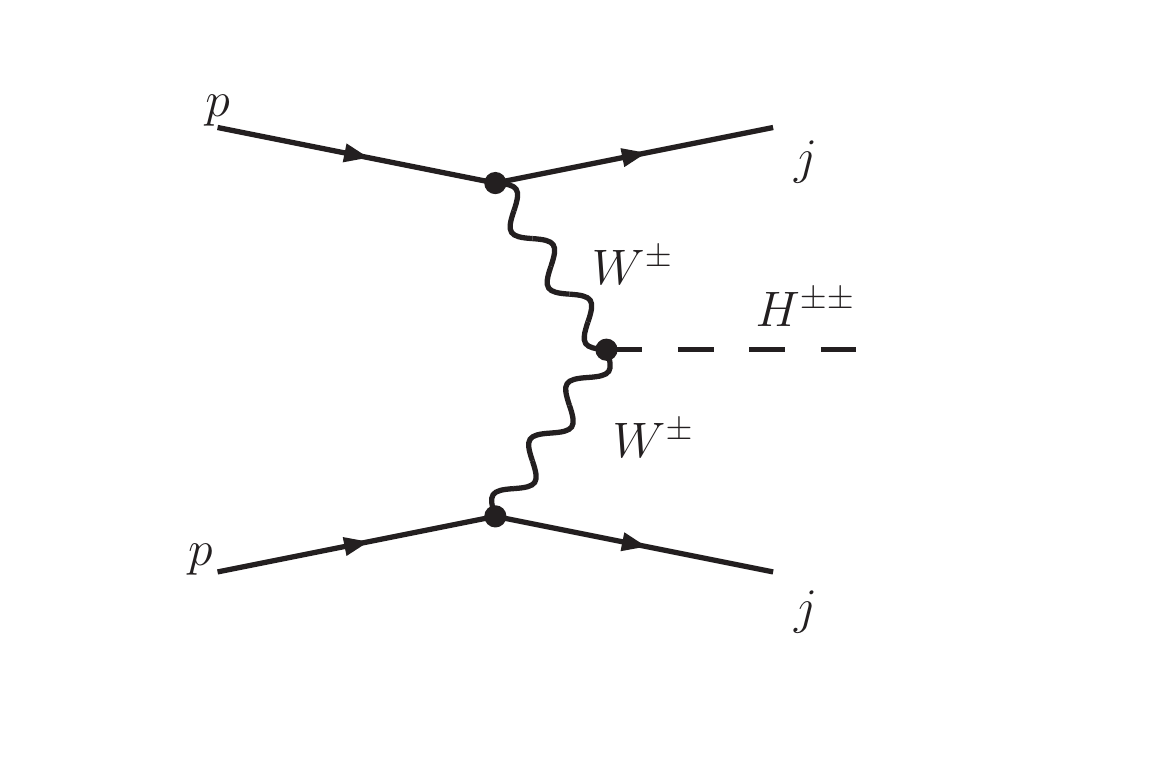} \\
\includegraphics[width=0.45\linewidth]{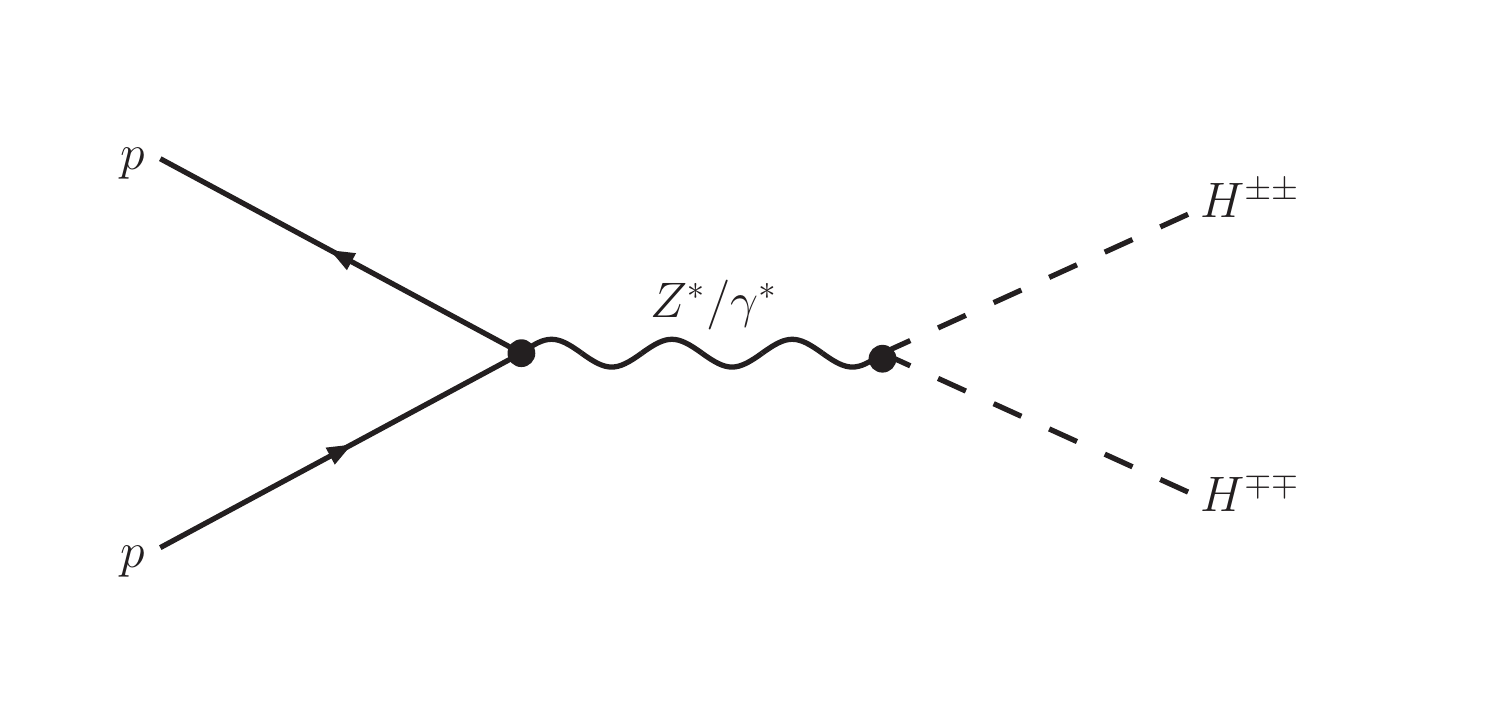}~~\includegraphics[width=0.3\linewidth]{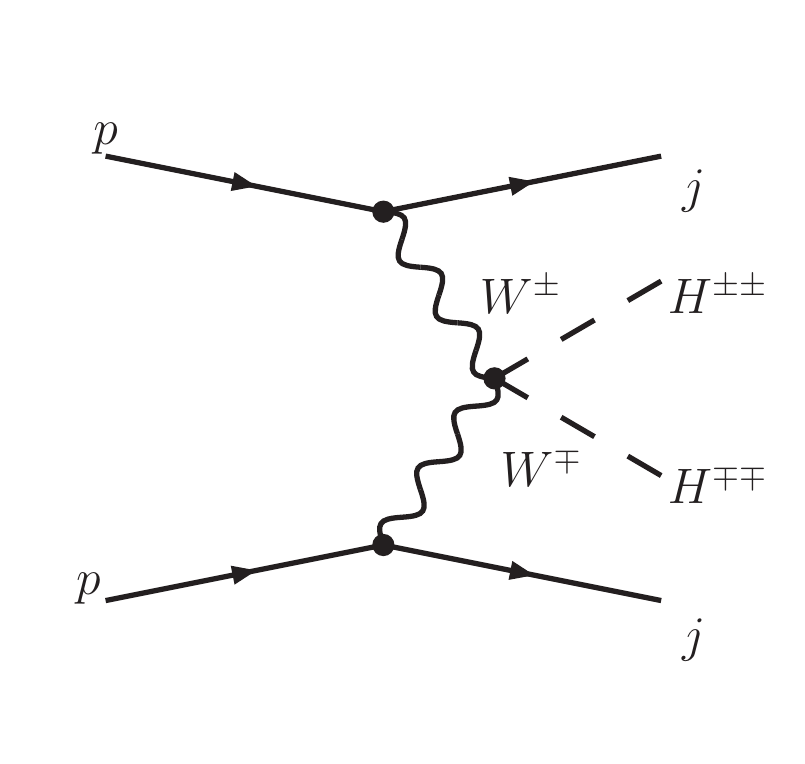}
\caption{Dominant Feynman diagrams for the production of doubly charged scalers $H^{\pm\pm}$ (i.e.\ the doubly charged components of the triplet Higgs field $\Delta$ in the minimal type II seesaw mechanism) via neutral and charged current interactions. }
\label{feynman} 
\end{figure}

In the following, we will focus on the doubly charged scalar $H^{\pm\pm}$, which has the clearest collider signatures.  
Under our assumptions (cf.\ section 2), it is the lightest of the new scalars and can decay to two same-sign leptons, $H^{\pm\pm}\to l^\pm_\alpha l^\pm_\beta$, to two on-shell W-bosons, $H^{\pm\pm}\to W^{\pm}W^{\pm}$ or into the three body final states $H^{\pm\pm}\to W^\pm (W^{\pm})^\ast\to W^\pm f\bar{f'}$, depending on the triplet VEV and the mass $m_{H^{\pm\pm}}$.
For $v_T < 10^{-4}$ GeV the decay to two same-sign leptons is dominant, cf.\ e.g.\ \cite{Dev:2013ff}. The production cross sections for all production modes of the triplet components are shown in fig.~\ref{cross_section} for $\sqrt{s}=13$~TeV and the example value $\upsilon_T = 0.1$ GeV, fixing  $\lambda'_{HT} = 0$ for illustration (such that $m_{H^{\pm}} = m_{H^{\pm\pm}}$). The corresponding Feynman diagrams are shown in fig.~\ref{feynman}. 

As one can see from fig.~\ref{cross_section}, the production cross section for the s-channel charged current process $pp\to W^{\pm}\to H^{\pm\pm} H^\mp$ is twice the production through the neutral current process $pp\to Z^*/\gamma^*\to H^{++} H^{--}$. In comparison, the t-channel production cross section is subdominant for small $m_{H^{\pm\pm}}$, but falls off less strongly with $m_{H^{\pm\pm}}$ such that $pp\to W^{\pm} W^{\mp} \to H^{++} H^{--}$ dominates above about 300 GeV. The t-channel production of a single $H^{\pm\pm}$ is suppressed by the triplet VEV (which in the plot is chosen as $v_T=0.1$).

We remark that, although we will focus on searches for doubly charged scalars, also the singly charged scalars are subject to LHC searches.
Here, due to the large backgrounds from single top, $t\bar t$, and multi-vector bosons, these searches are not as stringent compared to those for the doubly charged scalars, see e.g.\ ref.\ \cite{Aaboud:2018gjj} and references therein.

\subsection{Impact on the Higgs-to-diphoton rate}
The decay of the (SM-like) Higgs boson into two photons is introduced at the one-loop level in the SM, and it is dominated by the contribution from top quarks and the gauge bosons $W^\pm$. In the SM the contribution of $W^{\pm}$ is dominant, the contribution from top quarks is smaller and has opposite sign.  The contributions from the doubly and singly charged scalars 
%have same sign as the contribution from $W$ and 
are proportional to the couplings 
\begin{align}
&g_{hH^{++}H^{--}} \approx \frac{\upsilon^2}{m^2_{H^{\pm\pm}}} \lambda_{HT}\;, \;\;\;\;g_{hH^{+}H^{-}} \approx \frac{\upsilon^2}{m^2_{H^{\pm}}}(\lambda_{HT}  +  \tfrac{1}{2}\lambda'_{HT})\,,
\label{coupling}
\end{align}
where we neglected a suppressed dependency on 
%$\theta$, 
the mixing angle of the CP-even components from the doublet and triplet scalar fields, which is assumed to be small. 

The currently reported signal strength from CMS in terms of the SM prediction is given by $\mu = \frac{\sigma^{exp}(h\to\gamma\gamma)}{\sigma^{SM}(h\to\gamma\gamma)}= 1.1^{+0.32}_{-0.3}$ \cite{CMS:ril}, which limits the contribution from the doubly and singly charged scalars to be 
%$\sim$100\% 
less than 100\% of the SM predicted value.
There is a broad region of parameters $\lambda_{HT}$ and $\lambda'_{HT}$ where this is satisfied (cf.\ e.g.\ \cite{Arhrib:2011vc}).

\subsection{LHC searches for prompt $H^{\pm\pm}$ decays}

\paragraph{Searches for same-sign lepton pairs:}
At the LHC, searches for decays to same-sign leptons have been performed at center-of-mass energies $\sqrt{s}=7$~TeV, $8$~TeV and $13$~TeV  \cite{ATLAS:2012hi,CMS:2011sqa,ATLAS:2014kca,CMS:2016cpz,Aaboud:2017qph,CMS:2017pet}. 
For $m_{H^{\pm\pm}} > 300$ GeV, the strongest constraints stem from  the data sets with 36.1/fb at $\sqrt{s} = 13$~TeV for same-sign $ee$, $\mu\mu$, $e\mu$ pairs from decaying $H^{++}H^{--}$ pairs. In the following we use the bounds from the ATLAS analyses. Their negative search results put stringent bounds on the production cross section of the doubly charged Higgs bosons.
When $H^{\pm\pm}\to l^\pm_\alpha l^\pm_\beta$ is the dominant decay mode, i.e.\ as long as $Y_\Delta$ is not too small (or $v_T$ is below $\sim 10^{-4}$ GeV), the cross section depends only on $m_{H^{\pm\pm}}$, and values of $m_{H^{\pm\pm}}$ below about 620 GeV can be excluded.

It is important to stress that the analyses mentioned above require the $H^{\pm\pm}$ to decay promptly to three different modes, same-sign $ee$, $\mu\mu$ and $e\mu$. The most stringent constraint for $m_{H^{\pm\pm}} < 300$ GeV comes from the di-muon final state searches with $8$~TeV (e.g.\ from the ATLAS analysis in ref.~\cite{ATLAS:2014kca}), where the ``promptness'' condition is defined via the longitudinal impact parameter $z_0$ and the (transverse) impact parameter $d_0$ of the reconstructed track as
\begin{align}
&|z_0 \times \sin \theta|< 1 \text{ mm}\, \text{ and} \nonumber \\
&|d_0| < 0.2 \text{ mm}\,.
\label{eq:displacement}
\end{align}
When we apply the constraints on the cross section from prompt same-sign lepton pair searches where the $H^{\pm\pm}$ might be comparatively long-lived, we take only the fraction  of events into account which satisfy these ``promptness'' criteria. We will discuss this in detail in the next section.

\paragraph{Searches for same-sign $W$ pairs:}
In ref.~\cite{Aaboud:2018qcu} a search for pairs of $W$ bosons has been performed at ATLAS with 36.1/fb. Only the region where the $W$ decays are dominant and the $W$ bosons are on-shell has been considered.
No excess above the SM predictions has been found. This leads to an exclusion of the mass region where $m_{H^{\pm\pm}}$ lies between $200$ and $220$~GeV for $BR(H^{\pm\pm} \to W^\pm W^\pm) \sim 1$, which is satisfied for $v_T \gsim 3 \times 10^{-4}$~GeV.

\subsection{Signatures of long-lived $H^{\pm\pm}$}\label{sec:long-lived}
\begin{figure}
	\centering
	\includegraphics[width=0.55\textwidth]{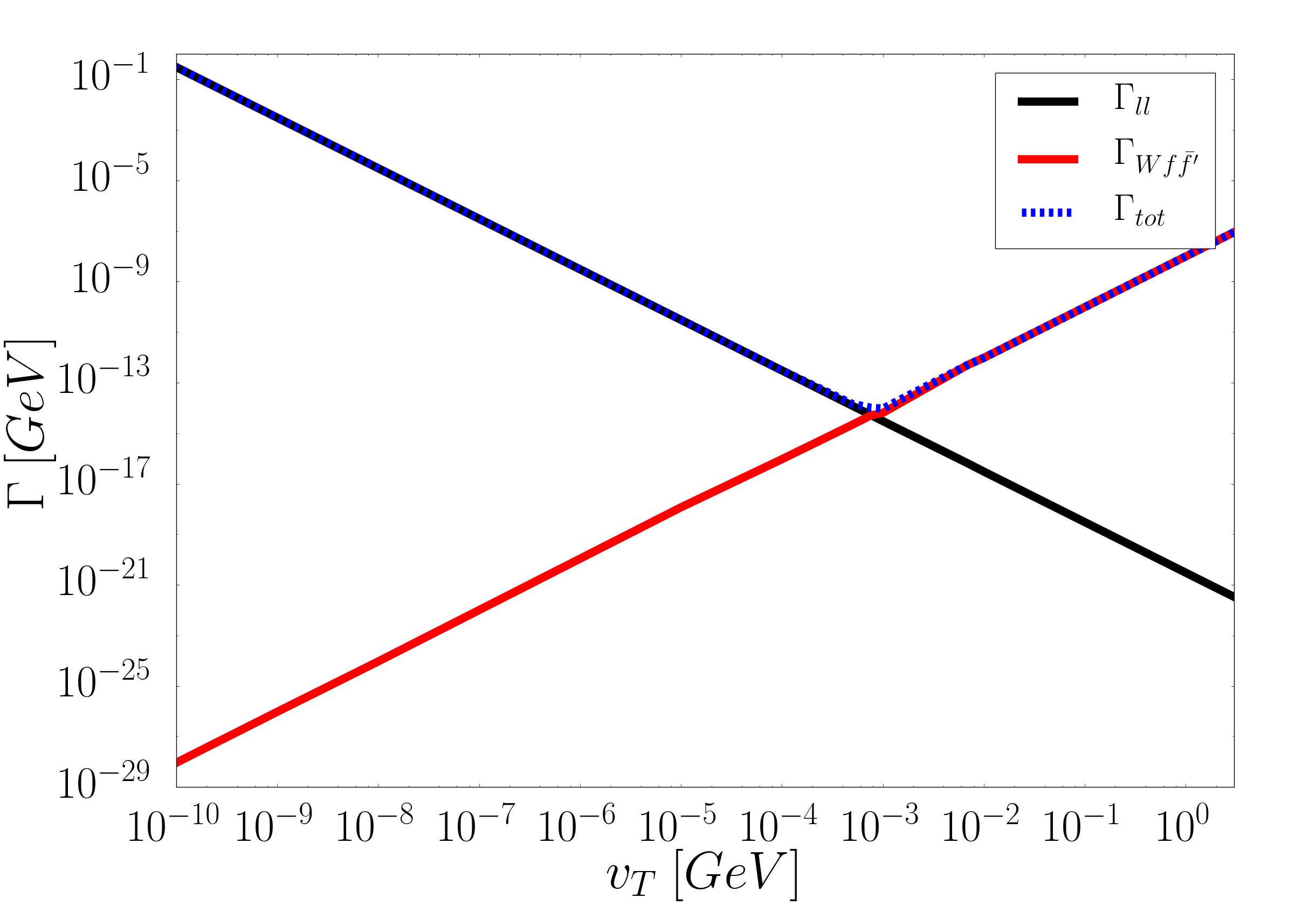}
	\caption{Total decay width of the doubly charged scalar field $H^{\pm\pm}$ as a function of the triplet VEV $v_T$ for $m_{H^{\pm\pm}}=130$~GeV (blue). Red and black lines are partial decay widths for $H^{\pm\pm}\to l^\pm l^\pm$ and $H^{\pm\pm}\to W^\pm (W^{\pm})^\ast\to W^\pm ff^\prime$ respectively.}
	\label{fig:partial}
\end{figure}

\paragraph{Lifetime of the doubly charged scalars at the LHC:}
For parameter values of the triplet VEV $v_T \lesssim 10^{-4}$ GeV, the decay of $H^{\pm\pm}$ into a pair of same-sign leptons is dominant (since  $Y_\Delta \propto 1/v_T$). For larger $v_T $ and the scalar mass $m_{H^{\pm\pm}} \lesssim 160$~GeV, the dominant decay to on-shell $W^\pm W^\pm$ is kinematically forbidden and the $H^{\pm\pm}$ decays mainly via $H^{\pm\pm} \to W^\pm (W^\pm)^* \to W^\pm f \bar f'$, where $f'$ is the isospin partner of the fermion $f$. 
The decay into a pair of same-sign leptons is proportional to $Y_\Delta$ and dominates for smaller value of  $v_T$. The rate of three body decays $H^{\pm\pm}\to W^\pm (W^\pm)^\ast \to W^\pm f \bar f'$ is proportional to $\upsilon_T$ \cite{kang:2014jia},
\begin{equation}
\Gamma(H^{\pm\pm}\to W^\pm (W^{\pm})^\ast \to W^\pm f \bar f^\prime) = \frac{g^6\upsilon^2_T m_{H^{\pm\pm}}}{6144\pi^3}\left(3+N_c\sum_{q,q^\prime}|V_{q,q^\prime}|^2\right)F\left(\frac{m^2_W}{m^2_{H^{\pm\pm}}} \right),
\end{equation}
with $N_c$ being the color factor and the factor of $3$ stems from the sum over the three lepton generations. The function $F(m^2_W/m^2_{H^{\pm\pm}})$ is given in Ref.~\cite{kang:2014jia}.
For the numerical analysis, we use the decay rate calculated with MadGraph \cite{Alwall:2014bza}.
Fig.~\ref{fig:partial} shows the total decay width (blue dotted line) as a function of $\upsilon_T$ for  $m_{H^{\pm\pm}} = 130$ GeV, where the red and black lines are the partial decay width for three body and same-sign di-leptons respectively. One can get a minimal total decay width (and hence a maximal lifetime) at the point where the two lines cross, which (for $m_{H^{\pm\pm}} = 130$ GeV) is at $\upsilon_T\sim 10^{-3}$ GeV.

\begin{figure}[!t]
\centering
\hspace{-0.5cm}\includegraphics[width=0.475\textwidth]{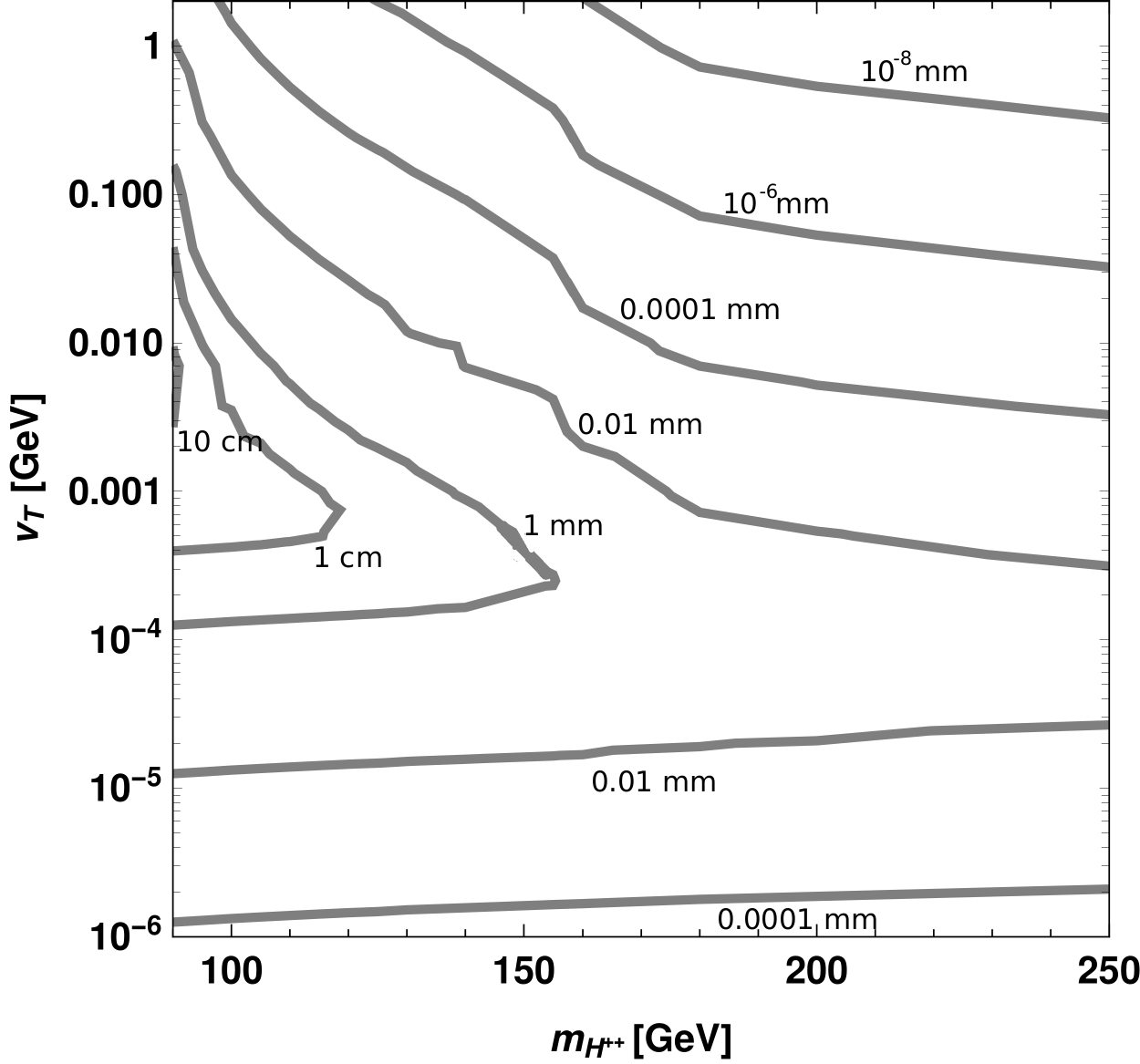}
\caption{Contours of proper lifetime of the doubly charged scalar particle $H^{\pm\pm}$ as a function of its mass and the triplet VEV $v_T$. }
\label{fig:lifetime}
\end{figure}

The resulting small total decay width gives rise to lifetimes for the $H^{\pm\pm}$ particles that can be macroscopic for certain parameter choices. We show the proper lifetime as a function of $m_{H^{\pm\pm}} $ and $v_T$ in fig.\ \ref{fig:lifetime}. It can be seen that between $v_T \sim 1 \times 10^{-4}$ GeV and $v_T \sim 1 \times 10^{-3}$~GeV and $m_{H^{\pm\pm}} < 155$~GeV a proper decay length above $1$~mm 
% can be obtained.
is possible.

\begin{figure}[!t]
\centering
\includegraphics[width=0.65\textwidth]{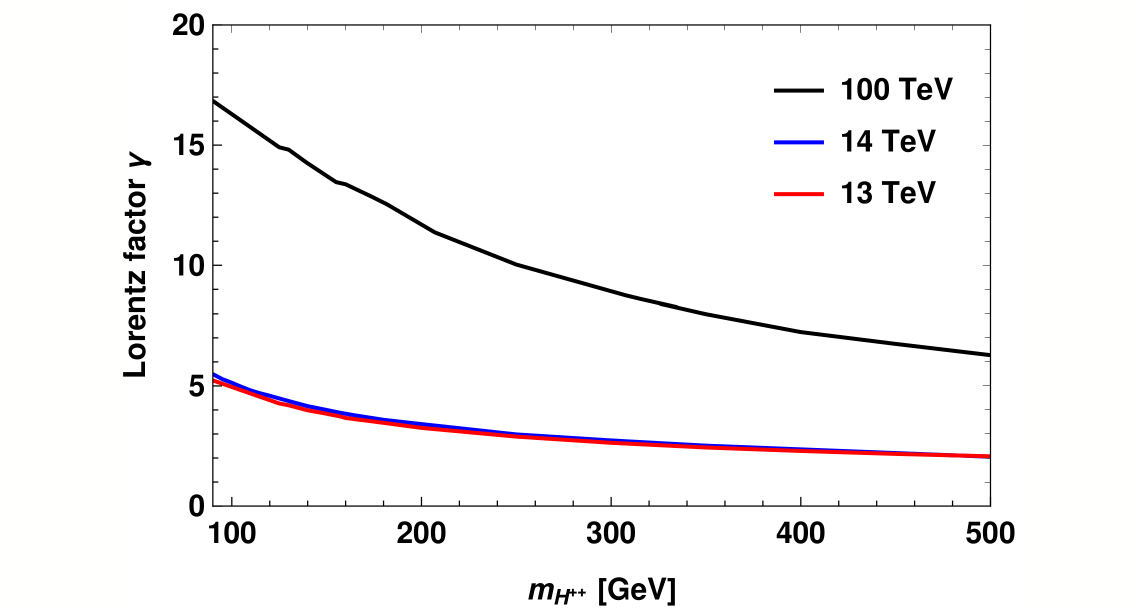}
\caption{Average Lorentz factor $\gamma$ as a function of $m_{H^{\pm\pm}}$ for $\sqrt{s} = 13$~TeV, $14$~TeV and $100$~TeV. }
\label{fig:distribution}
\end{figure}

\paragraph{Displaced vertex probabilities:}
The number of displaced $H^{\pm\pm}$ decays for a given parameter point can be expressed as:
\begin{align}
N(x_1,x_2,\sqrt{s},\mathcal{L}) = P(x_1,x_2)\sigma_{H^{\pm\pm}}(\sqrt{s})\mathcal{L},
\label{eq:numberofevents}
\end{align}
with $\sigma_{H^{\pm\pm}}(\sqrt{s})$ being the inclusive production cross section of a single $H^{\pm\pm}$, and $\mathcal{L}$ being the considered integrated luminosity.
$P(x_1,x_2)$ is the probability for a particle with a given proper lifetime $\tau$ to decay within given boundaries in the detector, defined by the range $x_1 \le x \le x_2$. It is given by: 
\begin{align}
P(x_1,x_2) = \int_{x_1}^{x_2}dx\frac{1}{c\tau \sqrt{\gamma^2 -1}}e^{(-\frac{x}{c\tau \sqrt{\gamma^2 -1}})} = e^{- \frac{x_1}{\Delta x_\mathrm{lab}}} - e^{- \frac{x_2}{\Delta x_\mathrm{lab}}}\,.
\label{probability}
\end{align}
where $\Delta x_\mathrm{lab}$ is the decay length in the laboratory frame given by (with the Lorentz factor $\gamma$)
\begin{align}
\Delta x_\mathrm{lab} = |\vec v|  \,\tau_\mathrm{lab} = c\tau \sqrt{\gamma^2 -1}\:,
\end{align}
and $\tau = \hbar/\Gamma$ with the total decay width $\Gamma$. For the Lorentz factor $\gamma$ of $H^{\pm\pm}$ we use average values  
obtained from simulations with MadGraph \cite{Alwall:2014hca}.
For the current LHC run at center-of-mass energy $13$~TeV, the HL-LHC at a center-of-mass energy 14 TeV, and for the FCC-hh with center-of-mass energy  $100$~TeV the average $\gamma $ is shown as a function of $m_{H^{\pm\pm}}$ in fig.\ \ref{fig:distribution}. 

\begin{figure}[!t]
\centering
\includegraphics[width=0.475\textwidth]{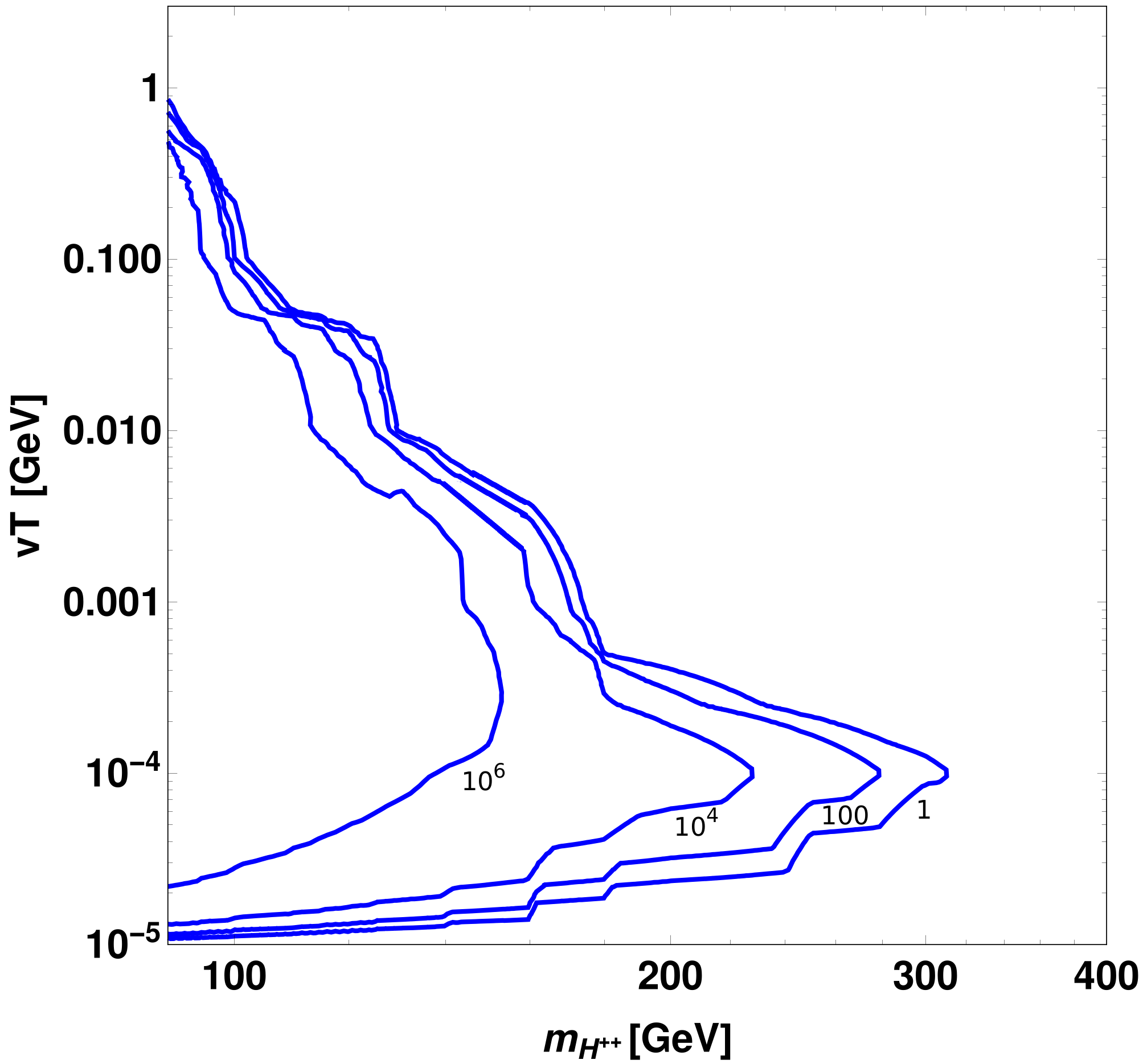}
\includegraphics[width=0.475\textwidth]{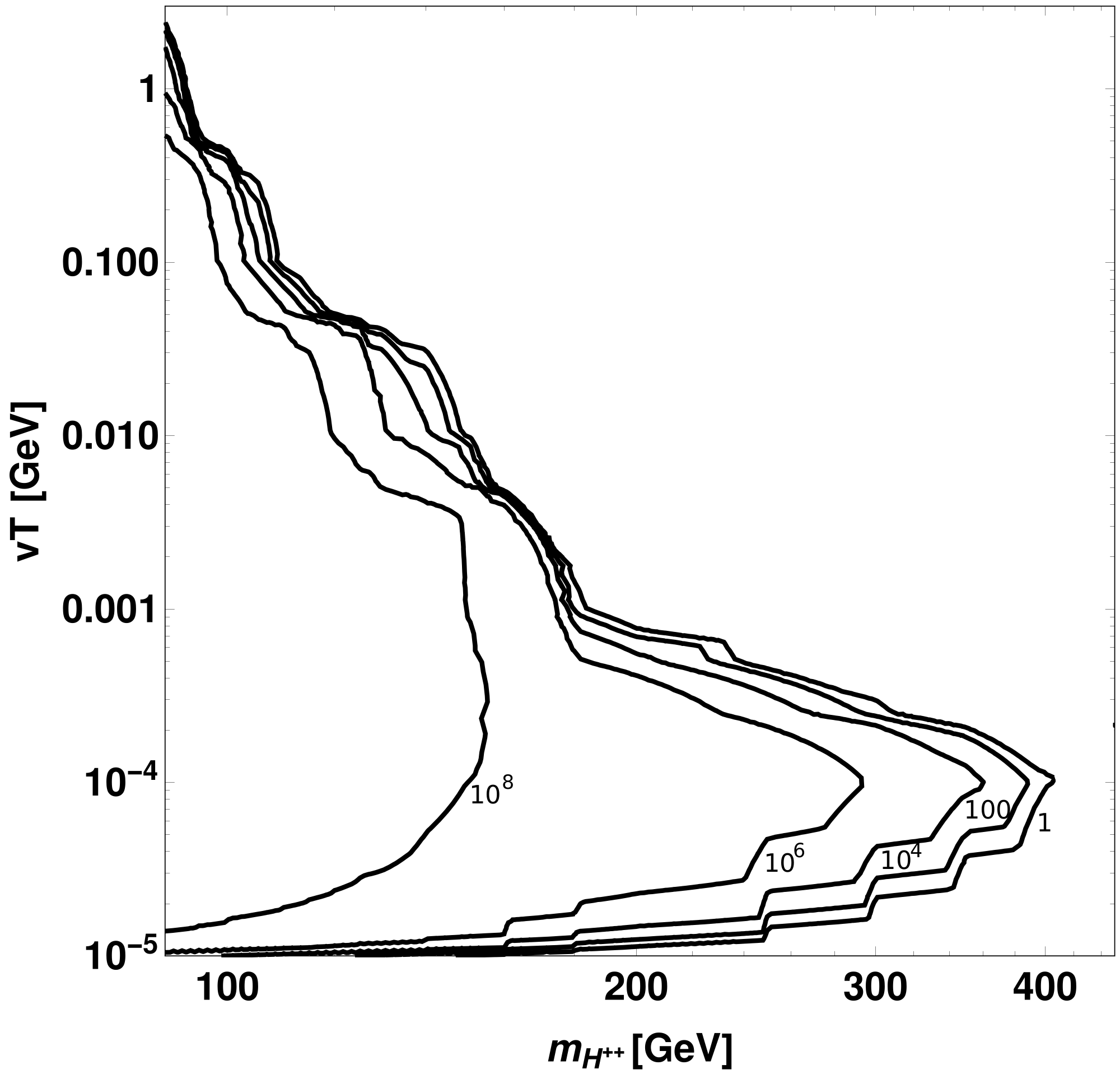}
\caption{Total number of doubly charged Higgs bosons decaying with a displacement between the boundaries $x_1 = 1$~mm and $x_2 = 1$~m, for the HL-LHC (left) and the FCC-hh (right). For this figure we consider the production channel $pp\to\gamma^\ast /Z^\ast\to H^{\pm\pm}H^{\mp\mp}$ only.}
\label{fig:numbers}
\end{figure}

For a first look at the prospects for displaced vertex searches, we consider the HL-LHC with $\sqrt{s} = 14$~TeV and the FCC-hh with $100$~TeV, and integrated luminosities of  $3000 \ fb^{-1}$ and $20 \ ab^{-1}$. We use eq.\ \eqref{eq:numberofevents} with the average Lorentz factors from fig.\ \ref{fig:distribution}, and the boundaries $x_1 = 1$ mm and  $x_2 = 1$ m. The numbers of displaced events are shown in fig.~\ref{fig:numbers} as a function of $m_{H^{\pm\pm}}$ and $v_T$. We remark that this first look is on the parton level and serves illustrative purposes only.

In the next section we will describe a possible LHC analysis to search for long lived doubly charged scalar bosons with $v_T$ = 5$\times 10^{-4}$ GeV and $m_{H^{\pm\pm}}$ = 130 GeV, where $c \tau  \approx 1$ cm. 
We will consider the pair production of doubly charged scalar through the neutral current $pp\to\gamma^\ast/ Z^\ast\to H^{\pm\pm}H^{\mp\mp}$ with two pairs of same sign di-lepton in the final state. 

We like to note that although the single production of a $H^{\pm\pm}$ in association with a single charged Higgs boson has a larger cross section (by a factor 2), it is not expected to significantly increase the prospects for a displaced vertex discovery. 
The reason is that the reconstruction of the single charged $H^\pm$ is not efficient since it decays mainly to a tau lepton and missing energy. We will therefore focus on the production channel $pp\to\gamma^\ast/ Z^\ast\to H^{\pm\pm}H^{\mp\mp}$.

\begin{figure}[!t]
\centering
\includegraphics[width=0.575\textwidth]{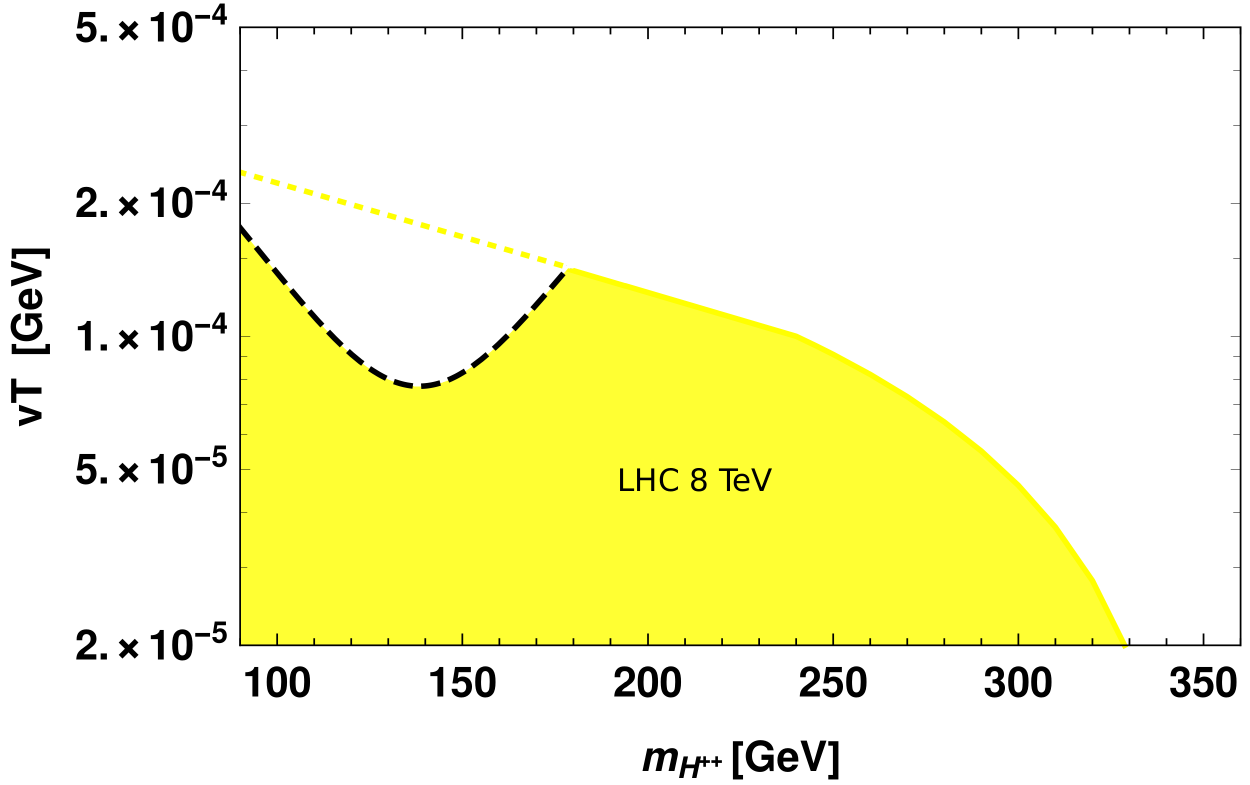}
\caption{Parameter space constraints from prompt LHC ($\sqrt{s}=8$~TeV) searches for same-sign dileptons at 95\% confidence level \cite{ATLAS:2014kca}, taking the possible displacement into account. The dashed black line indicates where the effective cross section is smaller than the observed limit. The dotted yellow line shows where the limit from the prompt search would be if all decays were prompt.}		
\label{fig:not_constraint_from_prompt}
\end{figure}

\paragraph{Application of constraints from prompt searches to potentially long-lived $H^{\pm\pm}$:}
As mentioned in the previous section, when applying the constraints on the $H^{\pm\pm}$ production cross section we have to take care that we only count the events where the ``promptness'' criteria of eq.~\ref{eq:displacement} are satisfied. We did this by simulating samples of events for the relevant parameter points to obtain the fraction of events which (for the given parameter point) satisfy the ``promptness'' criteria. This fraction is then multiplied with the total production cross section to obtain the ``effective'' production cross section to be compared with the constraints from the experimental analysis \cite{ATLAS:2014kca}.  To simulate samples for a wide range of parameter points, we performed a fast detector simulation using the same cuts as in \cite{ATLAS:2014kca}, and extracted $|z_0 \times \sin\theta|$ as well as the impact parameter $d_0$. 
The resulting excluded region from prompt searches for decays $H^{\pm\pm}\to l^\pm_\alpha l^\pm_\beta$ is shown in fig.~\ref{fig:not_constraint_from_prompt} as a function of $m_{H^{\pm\pm}}$ and $\upsilon_T$.

\paragraph{Searches for heavy stable charged particles at the LHC:}
Searches for heavy stable charged particles (HSCPs) have been performed by ATLAS (cf.\ e.g.\ \cite{Aad:2013pqd,Aad:2015oga}) and CMS (cf.\ e.g.\ \cite{Khachatryan:2016sfv}). They require that the HSCP candidate are stable on collider scales, i.e.\ they pass the relevant parts of the detector. For the ATLAS analysis, the HSCP candidate has to pass the muon system, while the CMS performed two versions of the analysis, one where the tracks have to pass the muon system, and a ``tracker only'' analysis where they only have to pass through the tracker (such that multiple hits in the tracker can be recorded). However, while the ATLAS analysis goes down to 50 GeV, the CMS analysis only starts at 100 GeV, and for HSCP candidates with $Q=2e$, they assume the candidate to be a lepton-like fermion (not a scalar as in our case). 
For a well reconstructed track the signature is a characteristic ionization energy loss ($dE/dx$).

To evaluate the constraint on the production cross section for $H^{\pm\pm}$ from HSCP searches, we must only count the events where the $H^{\pm\pm}$ actually pass through the relevant parts of the detector. This means, we have to use the  ``effective'' cross section $P(x_1,x_2) \sigma$ (cf.\ eq.~\eqref{probability}) with $x_1$ being the outer radius of the respective detector part, and $x_2=\infty$, i.e.\ the probability 
\begin{align}
P(x_1,\infty) = e^{- \frac{x_1}{\Delta x_\mathrm{lab}}} \:.
\end{align}   
For example, for $\gamma \sim 4 $, $m_{H^{\pm\pm}}=130$ GeV, $v_T = 5\times 10^{-4}$ GeV,  i.e.\ the benchmark point we will consider in the next section, we roughly get  $P(1\,\mbox{m},\infty)\sim 10^{-47}$ (for passing the tracker) and $P(11\,\mbox{m},\infty)\sim 10^{-182}$ (for passing the muon system). This clearly means that HSCP constraints cannot exclude this parameter point (in contrast to what has been claimed recently in \cite{Dev:2018kpa}). 
On the other hand, for $m_{H^{\pm\pm}} = 90$~GeV, $v_T = 7.5\times 10^{-4}$~GeV, where $c\tau \sim 35$~cm and $\gamma \sim 5$, one obtains $P(1\,\mbox{m},\infty) \sim 0.56$ and $P(11\,\mbox{m},\infty) \sim 10^{-3}$. Also this parameter point is not excluded by the ATLAS analysis which requires a track that passes the muon system, whereas a ``tracker only'' analysis (as performed by CMS) could quite likely exclude it. So far, however, this analysis does not exist for such low masses and for doubly charged scalars. It would therefore be highly desirable to extend the search to scalars with lower masses, and ideally also to the case of finite lifetimes.

Finally, we note that HSCPs can be searched for very well in the particularly clean environment of a lepton collider. At LEP, these searches have been done, cf.\ refs.\ \cite{Barate:1997dr,Abreu:2000tn,Abbiendi:2003yd} (cf.\ also ref.\ \cite{Abbiendi:2001cr} for prompt searches). They put stringent limits on the production cross section of heavy charged particles that manage to escape from the detector and exclude them for masses up to the kinematic limit of $\sim 90$ GeV. For finite lifetimes one may also reconsider these limits, however we expect that in the cleaner environment of a lepton collider a $H^{\pm\pm}$ with $m_{H^{\pm\pm}} \lesssim 90$ GeV would not have been missed. In the following, we will therefore focus on $H^{\pm\pm}$ masses above this value.

\begin{figure}%[!htb]
\centering
\includegraphics[width=\textwidth]{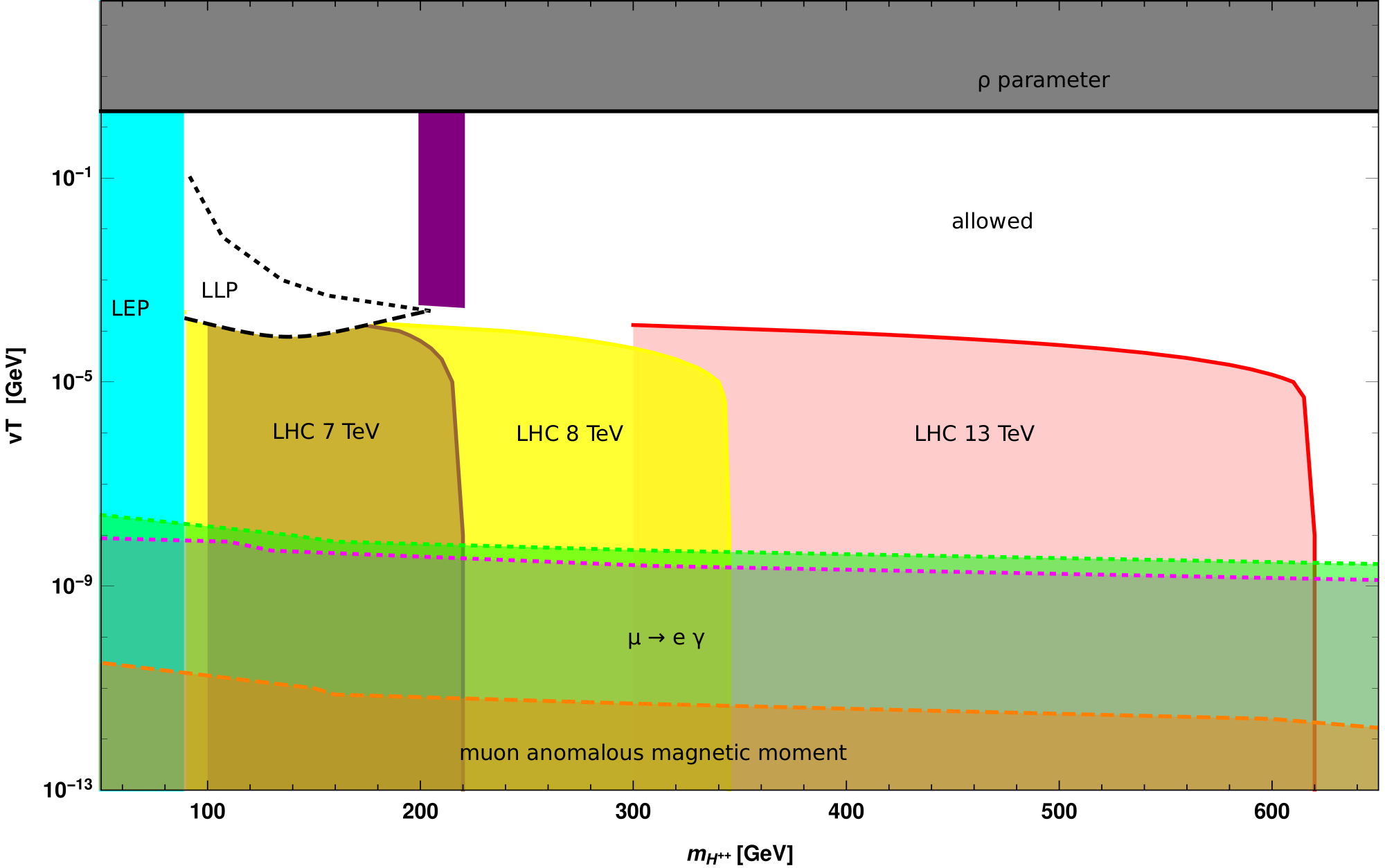}
\caption{Parameter space of the type-II seesaw model. The black area in top is excluded because of the $\rho$ parameter. The cyan vertical area is the estimate for the excluded region by searches at LEP. 
The orange region on the bottom is excluded by the experimental measurement for the muon anomalous magnetic moment. The magenta area is excluded by $\mu\to e\gamma$ (for our example choice of PMNS parameters and neutrino mass spectrum) and the green area is excluded by constraints on $\mu\to \bar{e}ee$. The red, yellow and brown areas are excluded by the LHC searches for same sign di-lepton final states at 7, 8 and 13 TeV. The purple area is excluded by LHC searches for same-sign $W$ bosons. Finally, the white area is allowed. The part of the white area inside the dashed and dotted black lines on the left (denoted by LLP) features displaced decays from long-lived $H^{\pm\pm}$. The lower dashed line is obtained from the limit on the prompt decays as described in the main text. The upper dotted line (where no experimental constraints exist to date) shows the region where $c\tau > 1$~mm. Above this line the dominant decay is the three-body decay to $W^\pm f \bar f'$. 
}
\label{fig:parameterspace}
\end{figure}

\section{Summary of present constraints}
We summarise the present constraints on doubly charged scalars $H^{\pm\pm}$ in the low scale type II seesaw scenario (under the simplifying assumptions discussed in section 2) in fig.\ \ref{fig:parameterspace}. The various constraints have been discussed in the previous sections. 
\begin{itemize}
\item We find that for $10^{-5} \text{ GeV} \lesssim v_T \lesssim 10^{-1}$ GeV and $m_{H^{\pm\pm}} \lesssim 200$ GeV there exists an allowed region where the $H^{\pm\pm}$ is long-lived and not excluded by neither prompt searches at LHC nor by the constraints from the existing HSCP analyses.
\item When the triplet vacuum expectation value is $v_T > 10^{-4} $ GeV, the decays $H^{\pm\pm}\to W^\pm W^\pm$ start to dominate the branching ratio, and the number of prompt decays $H^{\pm\pm}\to l^\pm_\alpha l^\pm_\beta$ is suppressed. Searches for di-W bosons are efficient only in the narrow range of $200 \, \mbox{GeV} \lesssim m_{H^{\pm\pm}} \lesssim 220  \, \mbox{GeV}$ \cite{Aaboud:2018qcu}, which is shown by the purple area in fig.\ \ref{fig:parameterspace}.
\item Finally, for $ m_{H^{\pm\pm}} \gtrsim 620$ GeV, constraints from LFV processes are the most powerful, constraining $v_T$ to be above about ${\cal O}(10^{-9})$ GeV for $ m_{H^{\pm\pm}} \sim 700$ GeV.   
\end{itemize}
It is striking that the part of parameter space where $v_T > 10^{-4} $ GeV is still largely untested by current experiments. However, this is the region where the low type II seesaw mechanism could be motivated by an approximate ``lepton number''-like symmetry. The symmetry would suppress the Yukawa couplings of the triplet to the lepton doublets and can thus provide a ``natural'' explanation for the smallness of the observed neutrino masses (in the t'Hooft sense that neutrino masses go to zero when the approximate symmetry is restored).\footnote{An alternative option consists in assigning lepton number to the triplet Higgs field. Then the parameter $\kappa$ for the coupling to the Higgs doublets would be suppressed by the approximate symmetry. This part of parameter space for the low type II seesaw mechanism is strongly constrained by LFV bounds.}  
Searches for displaced vertex signatures, as discussed in the next section, can help to probe part of this physically well-motivated parameter space.

\section{Displaced vertex signature: Analysis for a benchmark point}
To study in detail the prospect for displaced vertex searches from $H^{\pm\pm}$ decays, we perform an analysis at the reconstructed level. 
As benchmark point we consider $v_T$ = 5$\times 10^{-4}$ GeV and $m_{H^{\pm\pm}}$ = 130 GeV, and for definiteness $\lambda'_{HT} = 0$ and the other parameters fixed as discussed in section 2.
For this benchmark point with $c \tau  \approx 1$ cm, we consider the three different hadron colliders: the LHC with $13$ TeV center-of-mass energy and integrated luminosity $100 \ fb^{-1}$, the HL-LHC with $14$ TeV center-of-mass energy and integrated luminosity $3000 \ fb^{-1}$, and the FCC-hh with $100$ TeV center-of-mass energy and integrated luminosity $20 \ ab^{-1}$. 
For each of these colliders we generate a Monte Carlo event sample with $10^6$ events, using pileup events = 50 per vertex.
The Monte Carlo simulations of signal and background is carried out with the event generator MadGraph5 version 2.4.3~\cite{Alwall:2014hca}. 
For parton shower and hadronisation we use Pythia6~\cite{Sjostrand:2006za}, while the fast detector simulation is carried out by Delphes~\cite{deFavereau:2013fsa}.

\paragraph{Event reconstruction efficiency:}
For lifetimes as small as for the here considered benchmark point the $H^{\pm\pm}$ decays dominantly within the first (few) layers of the pixel tracker, and we consider the corresponding reconstruction efficiency to be equal to those of prompt signatures. We note that the track-only analysis is not sufficient to probe parameter points with such small lifetimes.

In general, for benchmark points with larger lifetimes the $H^{\pm\pm}$ decays may occur anywhere in the detector system, e.g.\ in the ECAL or in the muon system. 
The particle ID algorithms, which depend on the full detector information, are thus non-trivially affected by the displacement of each event.
Since our parent particle is electrically charged and has a very characteristic dE/dx we assume, however, that 100\% of its decays can be detected and identified, provided they are being caught by the triggers and the analysis selection requirements.

\paragraph{Selection requirements:}
For signal event selection we require at least one pair of charged tracks for the final state leptons, with lepton transverse momenta $P_T(\mu)> 25$ GeV and $|\eta(\mu)|<2.5$. We consider here only muons for simplicity, also in parts because it is not clear to us what kind of signal an electron would cause that appears inside the HCAL or muon system.
We use a muon isolation cone radius of $0.1$ and we impose a cut of $\Delta R>0.2$ between two same sign muons to ensure their separation. To increase the cut efficiency we impose further a cut on the invariant dimuon mass to be $M_{\mu\mu} = m_{H^{\pm\pm}} \pm 20$ GeV.

Furthermore, we require at least one displaced decay with same sign dimuons with a displacement in the $XY$ plane $L_{xy} > 8$ mm and the impact parameter $d_0>4$ mm.
This is expected to remove possible SM backgrounds and detector effects \cite{Cerdeno:2013oya,Aad:2012zx,Abdallah:2018gjj}. 
Finally, a matching condition between our reconstructed events and generator level events is imposed to ensure that the reconstructed tracks stem from the $H^{\pm\pm}$ candidate. Therefore we require the difference  $\Delta R (H^{\pm\pm}) $ of reconstructed and generator events to be $\Delta R (H^{\pm\pm})< 0.1$ \cite{Abdallah:2018gjj}.

\begin{figure}[h]
\includegraphics[width=0.52\textwidth]{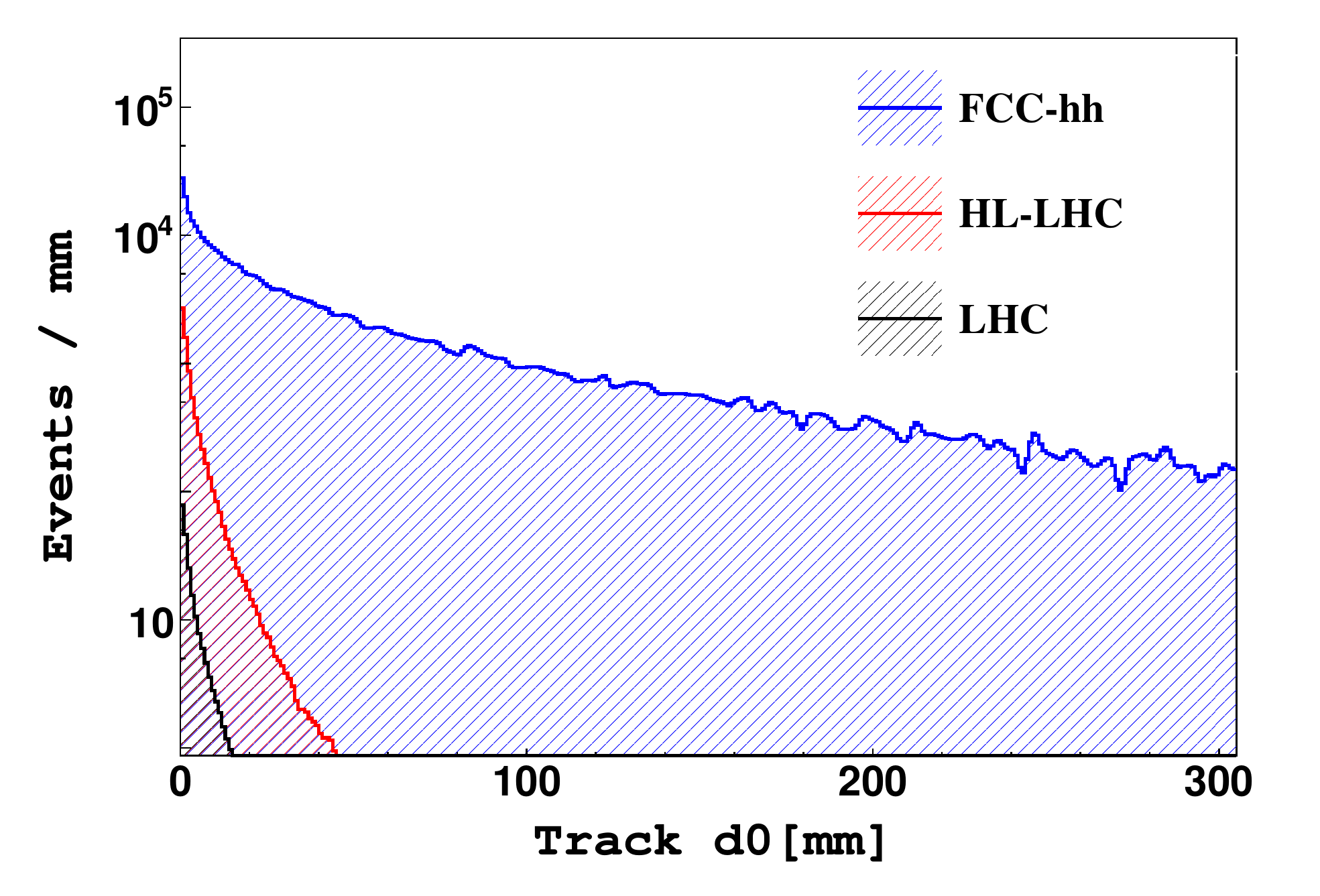}
\includegraphics[width=0.52\textwidth]{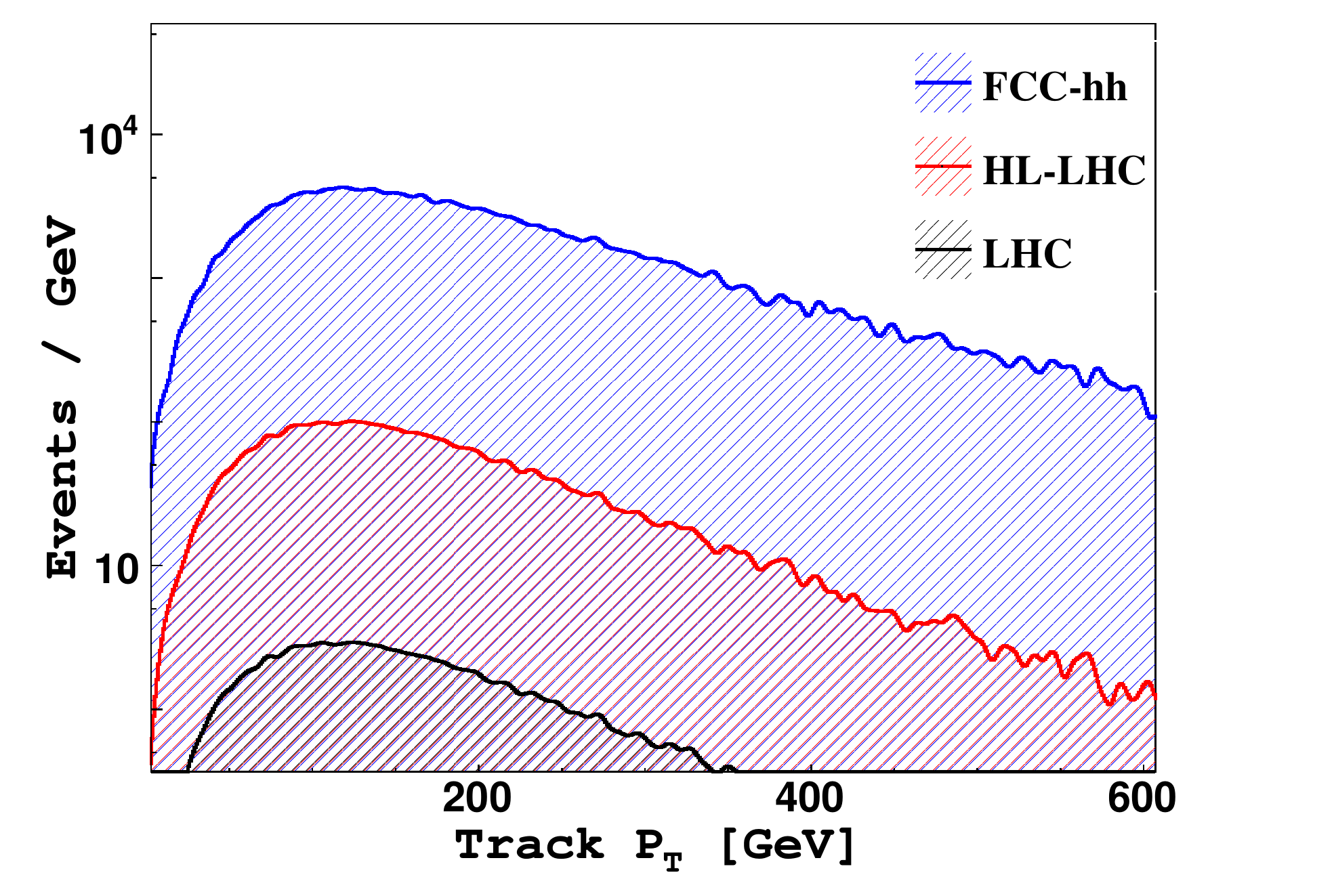}
\caption{Results from our simulations before applying any cuts. Left: impact parameter of the reconstructed track of $H^{\pm\pm}$ decaying to di-muons. Right: transverse momentum of the reconstructed track.}
\label{track_1} 
\includegraphics[width=0.52\textwidth]{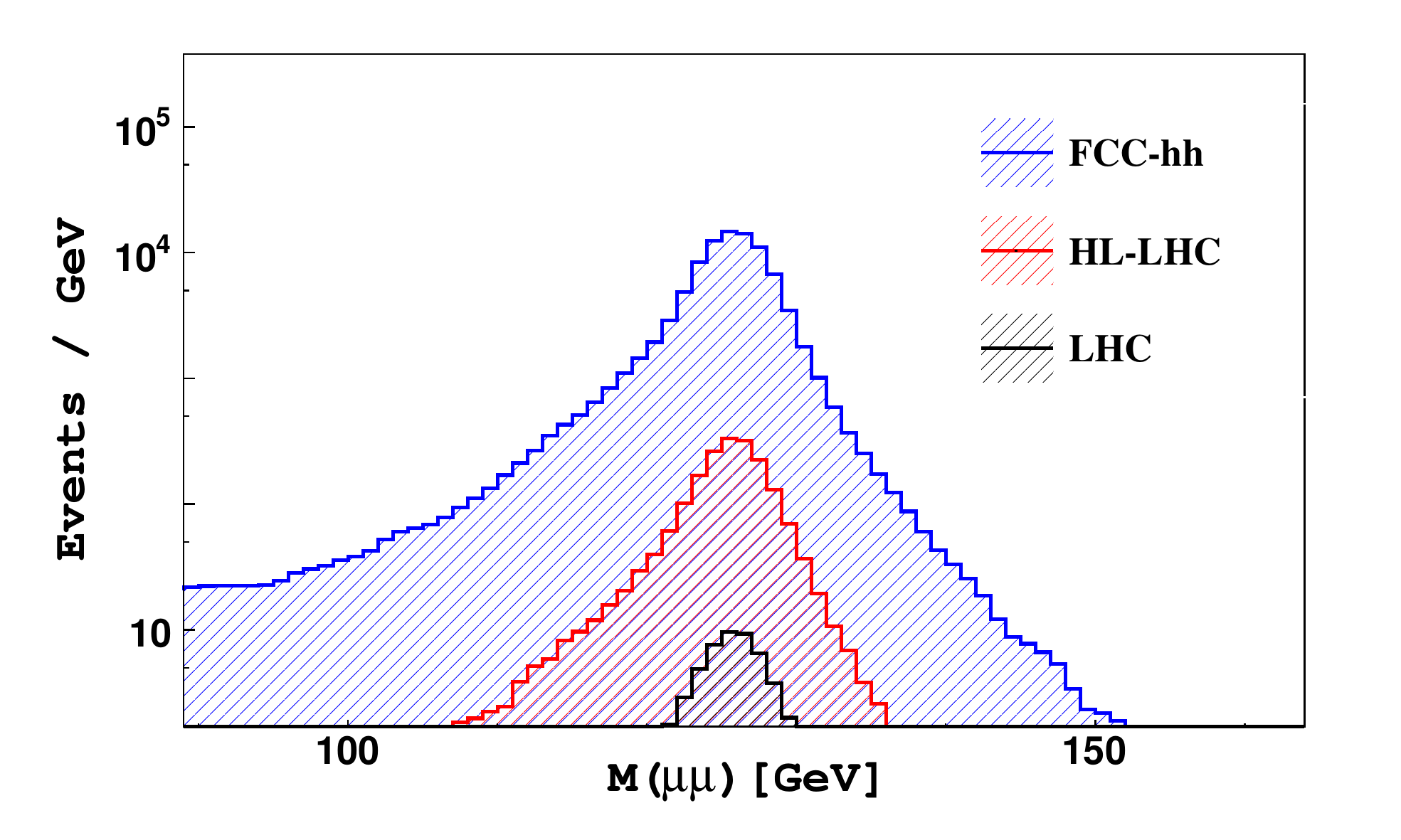}
\includegraphics[width=0.52\textwidth]{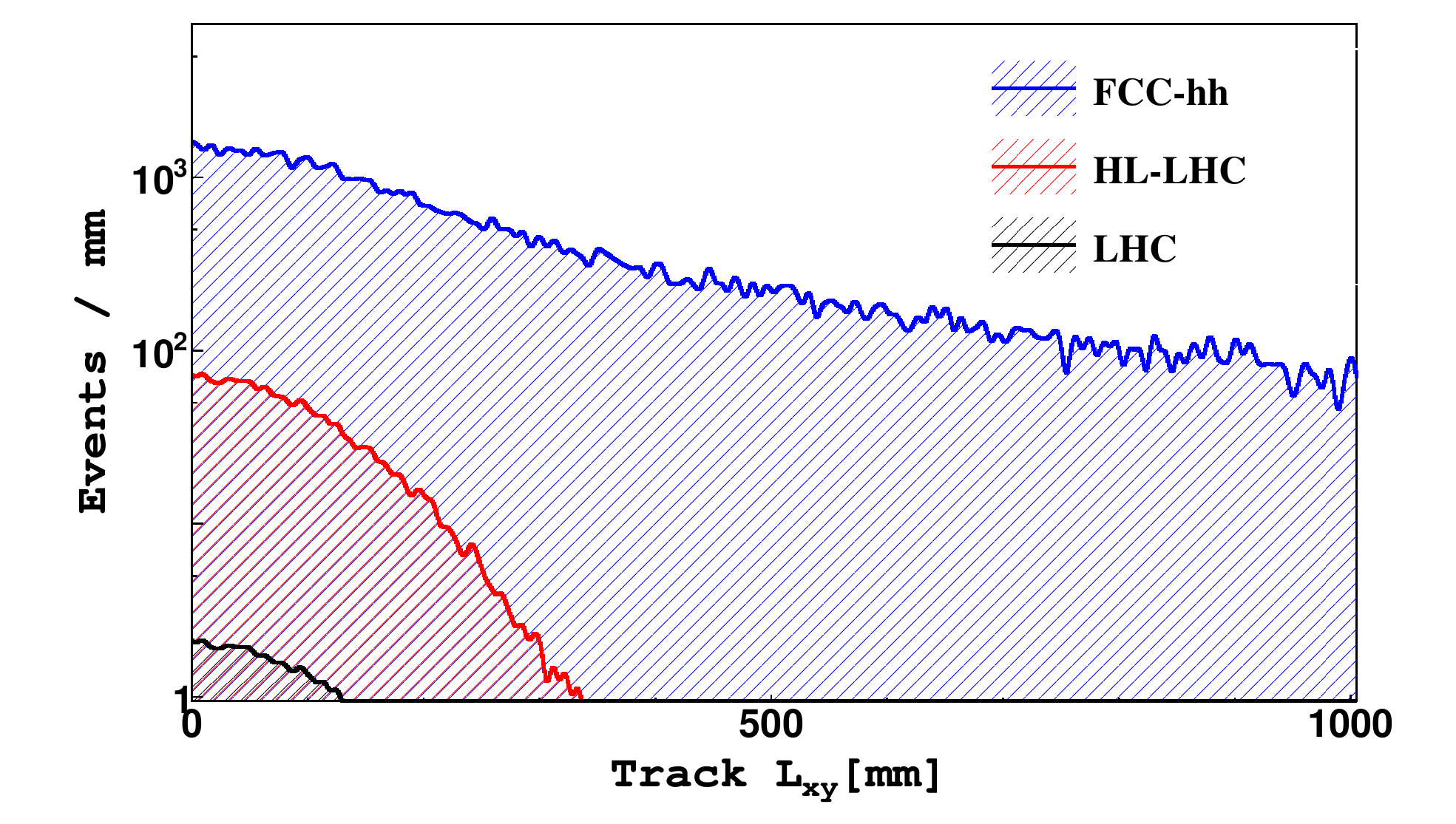}
\caption{Results from our simulations before applying any cuts. Left: invariant mass of $H^{\pm\pm}$ decaying to two muons final state. Right: longitudinal length  of $H^{\pm\pm}$ decaying to two muons.}
\label{track_2} 	
\end{figure}

\paragraph{Results:}
From the simulated event samples we reconstruct the $H^{\pm\pm}$ track and its displacement parameters from the observed distribution of the same-sign lepton pairs on an event-by-event basis.
Fig~ \ref{track_1} shows the resulting displacement of the secondary vertex (defined by the $H^{\pm\pm}$ decay) and the transverse momentum of the $H^{\pm\pm}$ candidate. In fig.~\ref{track_2} we show the invariant mass of the lepton pair (here two muons) and the transverse displacement of the secondary vertex.
All histograms are normalized to the expected number of events at the LHC, HL-LHC and FCC-hh, considering the corresponding integrated luminosity, before applying any cuts.

After applying the selection cuts, the cut flow of which is shown in tab.\ \ref{table:cut}, we find that about 13 events remain in the LHC data set, while for the HL-LHC and FCC-hh as many as $\sim 500$ and $\sim 32000$ events remain that are conform with our selection criteria. 
It is worth mentioning that, while the same benchmark point is used for different detector simulation and normalization factors (cross section$\times$ integrated luminosity), the detector dimensions as well the different value of the Lorentz factor $\gamma$ affect the analysis, greatly enhancing the number of signal events at the FCC-hh.

%%%%%%%%%%%%%%%%%%%%%%%
\begin{table}[ht]
\caption{Cut flow of simulated signal samples for displaced decays of the $H^{\pm\pm}$ to same sign dimuons. 
For this table, the benchmark point with $v_T$ = 5$\times 10^{-4}$ GeV and $m_{H^{\pm\pm}}$ = 130 GeV was considered. 
For the LHC, HL-LHC, and FCC-hh we use $13$, 14, and 100 TeV center-of-mass energy and an integrated luminosity of $100 \ fb^{-1}$, $3000 \ fb^{-1}$, and $20 \ ab^{-1}$, respectively. In our analysis we consider the production channel $pp\to\gamma^\ast Z^\ast\to H^{\pm\pm}H^{\mp\mp}$ only.}

\centering 
\begin{tabular}{c|c|c|c}
\hline\hline
Cuts & LHC & HL-LHC & FCC-hh \\ [0.5ex]
\hline\hline
 Expected events (detector level) & 280 & 10640 & 345323  \\
 \hline
 Two same sign muons  & 220 & 8135 & 244050 \\
 \hline
$P_T(\mu)>25 \text{ GeV} \& |\eta(\mu)|<$2.5$\&\Delta R(\mu,\mu) > 0.2 $ & 180 & 6508 & 209883  \\
\hline
110 GeV$< m_{H^{\pm\pm}}< $ 150 GeV & 175 & 6332 & 203586  \\
\hline
$L_{xy} > $ 8 mm & 76 & 2749 & 105864  \\
\hline
$d0 > $ 4 mm & 13.6 & 467 & 31759  \\
\hline
\end{tabular}
\label{table:cut}
\end{table}

\section{Conclusions}
In this paper we have investigated present constraints and displaced vertex signature prospects in the low scale type II seesaw mechanism, which is an attractive way to generate the observed light neutrino masses. It postulates a SU(2)$_\mathrm{L}$-triplet scalar field, which obtains an induced vacuum expectation value after electroweak symmetry breaking, giving masses to the neutrinos via its couplings to two lepton SU(2)$_\mathrm{L}$-doublets. 

Taking into account all relevant present constraints, including charged lepton flavour violation as well as collider searches, we have discussed the current allowed parameter space of the minimal low scale type II seesaw model. We investigated the possibility that the triplet components can be long lived, and  calculated carefully the constraints from the prompt searches, taking into account only the simulated events which satisfy the ``promptness'' criteria applied in the experimental analyses. 

We have also reconsidered constraints from present HSCP searches. 
We find that for most of the relevant parameter space for the long lived doubly charged scalars they cannot be applied because the lifetimes are not large enough to pass through the relevant parts of the detector. 
Nevertheless, such searches could test the part of the parameter space with lifetimes above a few cm via a ``tracker only'' analysis. 
Such analyses applicable to long lived doubly charged scalars do not exist but would be very desirable.

For $10^{-5} \text{ GeV} \lesssim v_T \lesssim 10^{-1}$ GeV and $m_{H^{\pm\pm}} \lesssim 200$ GeV, there exists an allowed region where the $H^{\pm\pm}$ is long-lived and not excluded by neither prompt searches at LHC nor by the constraints from the existing HSCP analyses.

For the characteristic displaced vertex signature where the doubly-charged component decays into same-sign charged leptons, we have performed a detailed analysis at the reconstructed level for a selected benchmark, which has a lifetime about 1 cm such that ``tracker only'' analyses are not efficient and additional information from secondary vertex reconstruction is necessary.
We found that already in present LHC data with 100 fb$^{-1}$ about 13 events may be detected in this way. 
Furthermore, the HL-LHC and FCC-hh have prospects to discover up to $\sim 500$ and $\sim 32000$ events in their final data sets, respectively.
Aside from the enhanced production cross sections and luminosities, the larger Lorentz factors at the FCC-hh/SppC \cite{Golling:2016gvc,Tang:2015qga} would lead to discovery prospects in an enlarged part of parameter space.

Finally, we like to point out that the symmetry protected low scale type II seesaw scenario, where an approximate ``lepton number''-like symmetry suppresses the Yukawa couplings of the triplet to the lepton doublets, is still largely untested by the current LHC results. Searches for displaced vertex signatures can help to probe part of this physically well-motivated parameter space.

\subsection*{Acknowledgements}
We are thankful to the organizers and participants the LHC LLP workshops for the stimulating atmosphere and many useful discussions. 
This work has been supported by the Swiss National Science Foundation.
O.F.\ received funding from the European Unions Horizon 2020 research and innovation program under the Marie Sklodowska-Curie grant agreement No 674896 (Elusives). 

\appendix

\vspace{0.7cm}

\bibliographystyle{unsrt}

\end{document}